\documentclass[aip,jmp,showpacs,showkeys,amsmath,amssymb,preprint]{revtex4-1}

\usepackage{graphicx}
\usepackage{epsfig}
\usepackage{bm}
\usepackage[dvips]{hyperref}
\def\beq{\begin{eqnarray}}
\def\eeq{\end{eqnarray}}
\def\nn{{\nonumber}}

\begin{document}


\title[]{Hamiltonian analysis for linearly acceleration-dependent Lagrangians}

\author{Miguel Cruz}
              \altaffiliation[On leave from:]{ Departamento de F\'\i sica, Centro de 
              Investigaci\'on y Estudios Avanzados del I.P.N., 
              Apdo. Postal 14-740, 07000, M\'exico D.F. M\'exico}
\affiliation{Facultad de F\'\i sica, Universidad Veracruzana, 91000, Xalapa, VER., M\'exico}

\email{miguelcruz02@uv.mx; roussjgc@gmail.com; efrojas@uv.mx} 

\author{Rosario G\'omez-Cort\'es}
\affiliation{Facultad de F\'\i sica, Universidad Veracruzana,
             91000, Xalapa, VER., M\'exico}

\author{Alberto Molgado}
              \altaffiliation[Also at ]{Dual CP Institute of High Energy Physics, M\'exico }
\affiliation{Facultad de Ciencias, Universidad Aut\'onoma de San Luis Potos\'{\i},
             Av. Salvador Nava S/N Zona Universitaria, CP 78290, San Luis Potos\'{\i}, SLP, M\'exico}
\email{molgado@fc.uaslp.mx}

\author{Efra\'\i n Rojas}
\affiliation{Facultad de F\'\i sica, Universidad Veracruzana,
             91000, Xalapa, VER., M\'exico}            

\begin{abstract}
We study the constrained Ostrogradski-Hamilton framework for the equations
of motion provided by mechanical systems described by second-order derivative 
actions with a linear dependence in the accelerations. We stress out the 
peculiar features provided by the surface terms arising for this type of 
theories and we discuss some important properties for this kind of actions 
in order to pave the way for the construction of a well defined quantum 
counterpart by means of canonical methods. In particular, we analyse in 
detail the constraint structure for these theories and its relation to the 
inherent conserved quantities where the associated energies together with a 
Noether charge may be identified. The constraint structure is fully analyzed
without the introduction of auxiliary variables, as proposed in recent works 
involving higher order Lagrangians.  Finally, we also provide some examples 
where our approach is explicitly applied, and emphasize the way in which our 
original arrangement results propitious for the Hamiltonian formulation of 
covariant field theories.
\end{abstract}



\maketitle

\date{\today}

\section{Introduction}
\label{sec:intro}

Gauge theories with higher derivatives have a potential interest for the study of 
the existence of several symmetries given their intrinsic nature. The Hamiltonian 
formalism provides a convenient way to formulate and explore the symmetries of 
field theories as well as to explore the ghost problem and the unitary properties 
of quantum gauge field theories for systems with higher-order singular Lagrangians. 
In particular, we are interested in linearly acceleration-dependent Lagrangians. 
Some interesting gauge theories presenting a linear dependence on acceleration, 
include General Relativity, cosmological brane models, and Electromagnetism with 
Chern-Simons corrections, for example, thus deserving special attention whenever the
quantum counterpart by means of canonical quantization procedures is aimed.
A common strategy to develop the Hamiltonian formalism for this type of theories 
is by means of the introduction of auxiliary variables which is accompanied by 
certain constraints enforcing the definition of the new variables. However, this 
procedure sometimes may become hard to manipulate since for a generic system we 
must work out the constraint algebra of an enlarged set of constraints. Thus, our 
standpoint will be to test the Ostrogradski-Hamiltonian (OH) formalism in order to 
completely analyze the constraint structure for the theories we are interested in, 
avoiding the cumbersome constraint analysis found in recent literature.

As mentioned before, in this paper we review the OH framework 
for Lagrangians with a linear dependence on the accelerations for which the Hessian vanishes 
identically~\cite{hojman1,hojman2,hojman3,Foster,udriste,barker,tesser,popescu1,popescu2}. 
Even though for this kind of Lagrangians one may find, under certain conditions, an equivalent 
Lagrangian where the acceleration terms go into a surface term, in this work we have opted 
to keep all the variables in the enlarged phase space in order to construct a natural Dirac 
algorithm to determine the physical constrained surface where dynamics will take place. 
Our motivation is thus to avoid the potential emergence of deviations in the 
Hamiltonian formalism induced by discarding surface terms associated to the general 
covariance of the system, as discussed for example in~\cite{Pirani}. In our opinion, this is 
an issue that has been overlooked in many contributions. In this sense, our study contributes 
to elucidate the relationship of the constraint analysis between the formalism for systems 
with boundary terms and Dirac's for first-order theories. We find that this constrained 
OH structure may be completely inferred from the associated Euler-Lagrange 
equations of motion. The approach considered here has been of particular relevance for 
the analysis of the canonical formalism  for singular systems for which the Hamiltonian 
does not result in a quadratic expression of the dynamical momenta involved.  Indeed, this 
is the case for certain brane models encountered either in the minisuperspace cosmological 
context or within Dirac's model of the electron as a charged 
membrane~\cite{ostro-prd,membrane,mgbg,ccmr}. Similar interpretations, as the ones carried 
out for the brane models, bring back into life the old idea that considers General Relativity 
as a second-order field theory~\cite{Dutt,Ghalati}. For a comparison, we encourage the reader 
to see the treatment of the Regge-Teitelboim model for General Relativity \`a la 
string either within the Hamiltonian formalism of an enlarged phase space established 
in~\cite{Paul,PaulRT}, or from the geometrical viewpoint related to tangent forms developed 
in~\cite{popescu1,popescu2}. 

Lagrangians of this type have been discussed extensively in the regular 
case~\cite{hojman1,hojman2,hojman3,hojman4,deleon,pardo,Gonera-note}. 
For the singular case, however, there are several points which seem to remain unclear and, 
therefore, this will be our purpose here: to analyse the singular nature of these Lagrangians 
by means of an OH approach. In particular, we resolve the conditions under which the system 
we consider allows surface terms and explore the role that this kind of terms play on the 
constraint structure of the theory. Further, we also emphasize the character of geometric 
invariants which provide well-defined energies and the general Noether charge, and examine 
their relevance for covariant systems through the so-called Zermelo 
conditions~\cite{Kondo,MironL,MironH,MironNoether,matsyuk,gracia,gracia2,dediego}.  
Although most of these last references are written in the elegant language of differential 
geometry, we decided to keep the widely known physicist notation in order to establish a 
direct comparison with recent works found in the literature where an analogous setup is 
considered for the analysis of the constrained structure for brane theoretical models by 
means of auxiliary variables.

This paper is organized as follows. In Section 2 we explain at some length why these 
Lagrangians are best understood straightforwardly as standard second-order Lagrangians 
instead of neglecting the surface term from the beginning. We compute both the general
energies and the Noether charge for this type of systems. Also, this Section serves 
as a guide in the transition towards the OH formalism for this type of theories. In 
Section 3 we describe the important role of the surface term, and we decompose our Lagrangian 
into two components, one related to the dynamic term, and the other related to the surface 
term. In Section 4 we complete the OH approach within Dirac formulation for constrained 
systems. Section 5 is dedicated to the geometric analysis of the constraints.  
We provide several examples of applications of our scheme in Section 6, related to the
chiral oscillator, the geometric dynamics resulting from a second-order 2-form and an 
optimal growth model from Economics, respectively.
In Section 7 we specialize our formulation to the analysis of covariant theories, and 
in particular, we develop our results for covariant brane theories 
where we present one further example given by a electrically charged bubble. 
In Section 8, we include some concluding remarks.  Finally, we address some technical issues 
related to conserved quantities and Helmholtz conditions in Appendices A and B, respectively.

\section{Lagrangians affine in acceleration}
\label{sec:reduced}

Let us consider the dynamical evolution of physical systems governed by the local action
\begin{equation}
S[x^\mu] = \int_c d\tau \, L(x^\mu,\dot{x}^\mu,\ddot{x}^\mu) ,
\label{eq:action}
\end{equation}
where the integral is well-defined for a parametric curve $c$, and the Lagrange function 
$L:T^2 M\rightarrow \mathbb{R}$ defined on a basis manifold $M$ is of the explicit form
\begin{equation}
 L(x^\mu,\dot{x}^\mu,\ddot{x}^\mu) = K_\mu (x^\nu,\dot{x}^\nu)\,\ddot{x}^\mu
+ V(x^\mu,\dot{x}^\mu)  .
\label{eq:lag1}
\end{equation}
Here, $K_\mu (x^\nu,\dot{x}^\nu)$ and $V(x^\mu,\dot{x}^\mu)$ are two arbitrary 
$\mathcal{C}^2(M)$ functions and $\mu,\nu = 0,1,2,\ldots, N-1$ label the local
coordinates on the extended configuration space. An overdot stands for a derivative 
with respect to a parameter $\tau$. We will consider for simplicity systems with a 
finite number of degrees of freedom. The Lagrangian (\ref{eq:lag1}) sometimes is referred 
to as {\it affine in acceleration} due to the fact that the  Hessian $H_{\mu \nu} = 
(\partial^2 L /\partial \ddot{x}^\mu \partial \ddot{x}^\nu)$ vanishes identically~\cite{udriste}.

As it is expected the Lagrangian $L$ is determined up to a total time derivative of the form 
\begin{equation}
 \widetilde{L}(x^\mu,\dot{x}^\mu,\ddot{x}^\mu) = L(x^\mu,\dot{x}^\mu,\ddot{x}^\mu)
+ \frac{d}{d\tau} Y  (x^\mu,\dot{x}^\mu) ,
\label{eq:lag2}
\end{equation}
that clearly do not affect the equations of motion as far as the function 
$Y(x^\mu,\dot{x}^\mu)$ is an arbitrary function not depending on accelerations. 
This arbitrariness 
imposes the following transformations 
laws for the functions 
$K_\mu$ and $V$
\begin{eqnarray}
\widetilde{K}_\mu (x^\nu,\dot{x}^\nu) &=& K_\mu + \frac{\partial Y}{\partial 
\dot{x}^\mu} ,
\label{eq:Ktransf}
\\
\widetilde{V} (x^\mu,\dot{x}^\mu) &=& V + \frac{\partial Y}{\partial {x}^\mu}
\dot{x}^\mu .
\label{eq:Vtrans}
\end{eqnarray}
These transformation laws closely resemble a sort of gauge transformation. Indeed, 
Eq.~(\ref{eq:Ktransf}) looks like an Abelian gauge transformation in the 
velocity coordinates sector whereas Eq.~(\ref{eq:Vtrans}) seems like an Abelian 
gauge transformation for a scalar potential in the position coordinates sector.  
Analogous transformations for Lagrangians affine in velocity were 
developed in~\cite{Govaerts}.

Associated with a generic second-order Lagrange function, $L$, the 
Euler-Lagrange (EL) equations of motion (eom) read 
\begin{equation}
E^{(0)}_\mu(L)=0 ,
\label{eq:eom1}
\end{equation}
where the differential operator $E^{(0)}_\mu$ is defined on the basis manifold as
\begin{equation}
\label{eq:operatorE0}
E^{(0)}_\mu:=\frac{\partial \ }{\partial x^\mu} - \frac{d}{d \tau} \left( \frac{\partial 
\ }{\partial \dot{x}^\mu} \right) + \frac{d^2}{d\tau^2} \left( \frac{\partial 
\ }{\partial \ddot{x}^\mu} \right) .
\end{equation} 
The $E^{(0)}_\mu$ operator is also known as the first Craig-Synge covector associated 
to a differentiable Lagrangian of second order (see~\ref{Noether} for details).

In our particular formulation, we have
\begin{eqnarray}
\frac{\partial L}{\partial x^\mu} &=& \frac{\partial K_\nu}{\partial x^\mu}\,
\ddot{x}^\nu + \frac{\partial V}{\partial x^\mu}  ,
\nonumber
\\
\frac{\partial L}{\partial \dot{x}^\mu} &=& \frac{\partial K_\nu}{\partial 
\dot{x}^\mu}\,\ddot{x}^\nu + \frac{\partial V}{\partial \dot{x}^\mu}  \,,
\nonumber
\\
 \frac{\partial L}{\partial \ddot{x}^\mu} &=& K_\mu  ,  
\nonumber
\end{eqnarray}
from which we obtain the equations of motion in the compact form
\begin{equation}
M_{\mu \nu} \ddot{x}^\nu = F_\mu ,
\label{eq:eom2} 
\end{equation}
provided that
\begin{equation}
\frac{\partial K_\mu }{\partial \dot{x}^\nu } =  \frac{\partial K_\nu }{\partial 
\dot{x}^\mu}.
\label{eq:id-1}
\end{equation}
This relation implies that $K_\mu$ should be related to a partial derivative of 
an arbitrary function with respect to the velocities $\dot{x}^\mu$, as may be inferred 
from transformation law~(\ref{eq:Ktransf}). We will explain in detail this last relation below.

Equations of motion in the form~(\ref{eq:eom2}) suggests that $M_{\mu \nu}$ may be 
interpreted as a ``mass-like matrix", while the term  $F_\mu$ may be interpreted as 
a ``force" vector. These quantities are defined as
\begin{eqnarray}
M_{\mu \nu} &:=& \frac{\partial P_\nu}{\partial x^\mu} - \frac{\partial 
p_\mu}{\partial \dot{x}^\nu},
\label{eq:Mmunu}
\\
F_\mu &:=&  \frac{\partial p_\mu}{\partial {x}^\nu}\,\dot{x}^\nu - 
\frac{\partial V}{\partial x^\mu} ,
\label{eq:force}
\end{eqnarray}
respectively. Here, we have introduced the quantities
\begin{eqnarray}
P_\mu &:=&  \frac{\partial L}{\partial \ddot{x}^\mu} = K_\mu, 
\label{eq:P1}
\\
p_\mu &:=& \frac{\partial L}{\partial \dot{x}^\mu} - \frac{d}{d \tau} \left( 
\frac{\partial L}{\partial \ddot{x}^\mu} \right) =
\frac{\partial V}{\partial \dot{x}^\mu} - \frac{\partial K_\mu}{\partial x^\nu}
\dot{x}^\nu ,
\label{eq:p1}
\end{eqnarray}
which are nothing but the canonical momenta in the so-called 
Ostrogradski-Hamiltonian approach (see 
Section~\ref{sec:ostro}). Conditions~(\ref{eq:id-1}) ensure that no 
third-order derivatives appear in the eom~(\ref{eq:eom1}). 
A trivial example of this is obtained by considering 
the particular case $K_\mu = K_\mu (x^\nu)$
as thus the eom are second order and conditions~(\ref{eq:id-1})
vanish identically. A different example is considered 
whenever $K_\mu$ specializes to 
$K_\mu = K_\mu (\dot{x}^\nu)$ for which the form~(\ref{eq:eom2}) is preserved
with $M_{\mu \nu} = - \partial p_\mu /\partial \dot{x}^\nu$. 
Thus, 
within our formulation, Eqs.~(\ref{eq:eom2}) are 
of second-order even if the action (\ref{eq:action}) involves second-order 
derivative terms. This issue is of a 
remarkable physical relevance since within this framework neither we have propagation 
of extra degrees of freedom nor we confront the instability issues associated to 
higher-order
theories~\cite{exorcist,nesterenko-insta,Stephen,llosa}.
This is so because condition (\ref{eq:id-1}) allows to split 
the original Lagrangian in two terms where one of them is a total time 
derivative which contains the acceleration dependence of 
the Lagrangian $L$. According to 
the dependence of $K_\mu$ and $V$, we have that $P_\mu = P_\mu (x^\nu,
\dot{x}^\nu)$ and $p_\mu = p_\mu (x^\nu,\dot{x}^\nu, P_\nu)$ and, in consequence 
$M_{\mu \nu}= M_{\mu \nu}(x^\alpha, \dot{x}^\alpha)$ and $F_\mu = F_\mu (x^\alpha,
\dot{x}^\alpha)$.

For a regular matrix $M_{\mu \nu}$, there is an inverse matrix $M^{-1} _{\mu \nu}$ 
such that $M^{-1} _{\mu \alpha} M_{\alpha \nu} = \delta_{\mu \nu}$ or $M_{\mu \alpha} 
M^{-1} _{\alpha \nu} = \delta_{\mu \nu}$. Hence, we can solve (\ref{eq:eom2}) 
for the accelerations $\ddot{x}^{\mu} = M^{-1\,\mu\nu} F_\nu$. 
This regular case is related to the so-called  inverse problem in 
Lagrangian mechanics which has been studied at length in \cite{hojman1,hojman2,hojman3,pardo}.
Clearly, we emphasize that to study the singular nature of the Lagrangian~(\ref{eq:lag1}) 
we must focus on the analysis of matrix $M_{\mu \nu}$. 

In addition, we may note that condition~(\ref{eq:id-1}) corresponds to an identically 
vanishing curl of $P_\mu$ in the velocity configuration space sector, namely
\begin{equation}
N_{\mu \nu} := \frac{\partial P_\nu}{\partial \dot{x}^\mu} -  \frac{\partial 
P_\mu}{\partial \dot{x}^\nu} = 0  \,. 
\end{equation}
It is straightforward to check that the quantities $M_{\mu \nu}$, $F_\mu$ 
and $N_{\mu \nu}$ are all invariant under the transformations~(\ref{eq:Ktransf}) 
and (\ref{eq:Vtrans}). Hence, it is not surprising that all these quantities 
will play a fundamental role in the following sections.

\subsection{Noether charge}

We describe now the conserved quantities for this type of theories. Our strategy will follow 
closely the content of Refs.~\cite{MironH,MironNoether}
~(see~\ref{Noether}, for further details). We begin
from Eqs.~(\ref{eq:GenEnergy1}) and (\ref{eq:GenEnergy2}) for the energies of a
second-order system associated to the vector field $W^\mu$ in terms of the 
momenta~(\ref{eq:P1}) and (\ref{eq:p1})
\begin{eqnarray}
\label{eq:energy1}
\mathcal{E}_c^{(1)}(L) & := &  - W^\mu P_\mu,
\\
\label{eq:energy2}
\mathcal{E}^{(2)}_c (L)& := & W^\mu p_\mu + \dot{W}^\mu P_\mu - L.
\end{eqnarray} 
We must emphasize that these energies are only conserved along the solutions of the 
so-called Craig-Synge covectors, as described in~\ref{Noether}, 
where we also show the way in which these energies are 
helpful in order to construct a Noether theorem for 
the Lagrangian we are working with.
With regards the function~(\ref{eq:QQ}) we have
\begin{equation}
\mathcal{Q}(L,\phi) = W^\mu p_\mu + \dot{W}^\mu P_\mu - 
\eta \,\mathcal{E}_c^{(2)}(L)+\dot{\eta} \,\mathcal{E}_c^{(1)}(L)-\phi ,
\label{eq:Noether}
\end{equation}
where $\mathcal{E}_c^{(1)}(L)$ and $\mathcal{E}_c^{(2)}(L)$ are given
by (\ref{eq:energy1}) and (\ref{eq:energy2}), and again we have used the 
definitions (\ref{eq:P1}) and (\ref{eq:p1}).
This function is conserved through an infinitesimal transformation of the form
\begin{equation}
\begin{split}
 x^\mu & \mapsto  x^\mu + \epsilon W^\mu (x,\tau) , \\
\tau  & \mapsto  \tau + \epsilon \eta(x,\tau)  ,
\end{split}
 \label{eq:infinitesimal}
\end{equation}
along the solution curves of the EL equations~(\ref{eq:eom1}).
In relations~(\ref{eq:Noether}) and~(\ref{eq:infinitesimal}) we have introduced
the differentiable vector field $W^\mu$ locally defined 
along a trajectory $c$ and such that it vanishes at the endpoints of that curve, 
the sufficiently small real number $\epsilon>0$, and the 
smooth arbitrary (locally defined) functions $\eta:=\eta(x,\tau)$ and 
$\phi:=\phi(x,\dot{x})$.   

We will realize in the following that for the case $W^\mu = 
\dot{x}^\mu$ the energies (\ref{eq:energy1}) and (\ref{eq:energy2}) become
\beq
\mathcal{E}_c^{(1)}(L) &=& - P_\mu \dot{x}^\mu,
\label{eq:energy3}
\\
\mathcal{E}_c^{(2)}(L) &=& p_\mu \dot{x}^\mu + P_\mu \ddot{x}^\mu - L.
\label{eq:energy4}
\eeq
The first one corresponds to the term leading to a total time derivative in the 
Lagrangian~(\ref{eq:lag1}), while the second one corresponds to the canonical Hamiltonian, 
$H_0$ (see Eq. (\ref{eq:H0}) below). For covariant brane theories these constitute 
first-class constraints generating gauge transformations~(see Section \ref{sec:covariant-brane}).
Also, for this particular case,  note that energies $\mathcal{E}_c^{(1)}(L)$ and 
$\mathcal{E}_c^{(2)}(L)$ are related trough the identity $\mathcal{E}_c^{(2)}(L)
=(\partial V/\partial\dot{x}^\mu)\dot{x}^\mu - V + (\partial \mathcal{E}_c^{(1)}(L)/\partial x^\mu)
\dot{x}^\mu$ as may be deduced from general expressions~(\ref{eq:GenEnergy1}) and~(\ref{eq:GenEnergy2}). 
 
In addition, on physical grounds, the function~(\ref{eq:Noether}) is nothing but the 
Noether charge associated to the action~(\ref{eq:action}), which may be expressed in terms of the 
momenta~(\ref{eq:P1}) and~(\ref{eq:p1}) as 
\begin{equation}
\mathcal{Q}(L,\phi)= - \eta \, H_0 + p_\mu \dot{x}^\mu + P_\mu \ddot{x}^\mu 
- [ \dot{\eta}\,P_\mu \dot{x}^\mu + \phi ],
\label{eq:Q-noether}
\end{equation}
where $W^\mu$ has been specialized to the velocity vector $\dot{x}^\mu$,
${\cal E}_c ^{(1)} = - \dot{x}^\mu P_\mu$ and ${\cal E}_c ^{(2)} = H_0 = 
p_\mu \dot{x}^\mu - V$ 
is the canonical Hamiltonian. 
It is worth noting that the results given by~(\ref{eq:energy1}), (\ref{eq:energy2})
and (\ref{eq:Noether}) are quite general for a second-order Lagrangian and comprise
a natural extension of the usual results for the Lagrangians $L(x^\mu,\dot{x}^\mu)$
to second-order Lagrangians $L(x^\mu,\dot{x}^\mu,\ddot{x}^\mu)$.

\section{On the surface term and s-equivalent Lagrangians}
\label{sec:surface}

We turn now to explore the conditions on $K_\mu$  such that partial 
integrations of the linear term in the accelerations in (\ref{eq:lag1})
does not lead to total time derivatives. If so, the condition $d(\dot{x}^\mu K_\mu)/d\tau
= 0$ is equivalent to
\begin{equation}
\left( K_\mu + \frac{\partial K_\nu}{\partial \dot{x}^\mu}\dot{x}^\nu\right)
\ddot{x}^\mu = -  \frac{\partial K_\mu}{\partial {x}^\nu}\dot{x}^\mu \dot{x}^\nu.
\label{eq:condition}
\end{equation} 
On the contrary, suppose now that we can identify a total time derivative. Without 
loss of generality we may define alternative quantities and rewrite the 
Lagrangian~(\ref{eq:lag1}) as follows
\begin{equation}
L(x^\mu,\dot{x}^\mu,\ddot{x}^\mu) = f (x^\mu,\dot{x}^\mu) +
\frac{d}{d \tau} \left[ g(x^\mu) h (\dot{x}^\mu) \right] , 
\label{eq:lag4}
\end{equation}
where $f(x^\mu,\dot{x}^\mu)$, $g(x^\mu)$ and $h(\dot{x}^\mu)$ are smooth functions.
In fact, the function $h (\dot{x}^\mu)$ is defined up to an integration constant. 
This results as a consequence that Eq.~(\ref{eq:id-1}) implies the 
existence of a boundary function $\Lambda (x^\mu, \dot{x}^\mu)$ such that $K_\mu=
\partial \Lambda / \partial \dot{x}^\mu$. For simplicity, from now on, 
we consider separation of variables for the boundary term.  From a more general viewpoint, 
analogous conclusions follow if we adopt the boundary function $\Lambda (x^\mu, \dot{x}^\mu)$ 
instead of the functions $g(x^\mu)$ and  $h (\dot{x}^\mu)$ \cite{querella}. 
(see also~\cite{hojman1}).

By expanding the total time derivative and comparing to Eq.~(\ref{eq:lag1})
yields
\begin{eqnarray}
K_\mu (x^\nu, \dot{x}^\nu ) &=& g\,\frac{\partial h}{\partial \dot{x}^\mu},
\label{eq:Kmu2}
\\
V(x^\nu, \dot{x}^\nu) &=& f + \dot{x}^\mu\,\frac{\partial g}{\partial x^\mu}\,h.
\label{eq:V2}
\end{eqnarray}
Expression (\ref{eq:Kmu2}) automatically fulfills condition (\ref{eq:id-1}), 
indeed. Hence, we can rewrite the original Lagrangian (\ref{eq:lag1}) as the 
sum of two Lagrangian functions
\begin{eqnarray}
{L}_d &:=& f = V - \frac{\partial g}{\partial x^\mu}\,h\,\dot{x}^\mu ,
\label{eq:lag-d}
\\
{L}_s &:=& \frac{d \left( g\, h \right)}{d \tau}
= g\,\frac{\partial h}{\partial \dot{x}^\mu}\,\ddot{x}^\mu
+ \frac{\partial g}{\partial x^\mu}\,h\,\dot{x}^\mu.
\label{eq:lag-s}
\end{eqnarray} 
Clearly, $L_d = L_d (x^\mu, \dot{x}^\mu)$ while $L_s = L_s (x^\mu, 
\dot{x}^\mu, \ddot{x}^\mu)$. The Lagrangian $L_s$ contains 
all the information related to the second-order derivatives. 
We will refer to these Lagrangian functions as the ``dynamic" 
Lagrangian and the ``surface" Lagrangian, respectively.
Lagrangians (\ref{eq:lag1})
and (\ref{eq:lag-d}), belonging  to the set known as {\it $s$-equivalent 
Lagrangians}, are characterized by eom which are solved by 
the same orbits~\cite{hojman1,hojman4}. 
The advantage of this separation is particularly conspicuous in certain 
cases of covariant theories. Indeed, sometimes $f(x^\mu,\dot{x}^\mu)$, 
regarded as an ordinary first-order Lagrangian, does not allow for a 
consistent Hamiltonian treatment whenever $M_{\mu \nu}$ is singular
(see, for example,~\cite{ostro-prd,membrane,davidson} as 
the Hamiltonian is not a simple quadratic expression in the momenta.
In addition, this separation helps to understand deeply the role of $M_{\mu \nu}$.

With respect to the surface Lagrangian, $L_s$, and considering the definitions
(\ref{eq:P1}) and (\ref{eq:p1}), we have the momenta
\begin{eqnarray}
P_\mu &=& g\,\frac{\partial h}{\partial \dot{x}^\mu},
\label{eq:P11}
\\
\mathfrak{p}_{\mu} &=& \frac{\partial g }{\partial x^\mu}\,h.
\label{eq:p11}
\end{eqnarray}
Analogously, with respect to the 
dynamic Lagrangian, $L_d$, we have the momenta
\begin{equation}
\mathbf{p}_{\mu} = \frac{\partial f}{\partial \dot{x}^\mu} =
\frac{\partial V}{\partial \dot{x}^\mu} - \frac{\partial g }{\partial 
x^\nu} \frac{\partial h }{\partial \dot{x}^\mu}\,\dot{x}^\nu 
- \frac{\partial g }{\partial x^\mu}\,h.
\end{equation}
Obviously, we have that with respect to the original Lagrangian (\ref{eq:lag1})
the conjugate momenta  to the position variables (\ref{eq:p1}) are given by
\begin{equation}
p_\mu = \mathbf{p}_{\mu} + \mathfrak{p}_{\mu} = \frac{\partial V}{\partial \dot{x}^\mu} - 
\frac{\partial g }{\partial x^\nu} \frac{\partial h }{\partial \dot{x}^\mu}\,
\dot{x}^\nu ,
\label{eq:pp1}
\end{equation}
that is, the total momenta $p_\mu$ associated to the Lagrangian~(\ref{eq:lag1}) 
(or (\ref{eq:lag4})) is obtained by adding a contribution from both the 
surface and the dynamic momenta. At this point it is clear what to do in 
order to express the original Lagrangian (\ref{eq:lag1}) in the fashion~(\ref{eq:lag4}). 
Starting from (\ref{eq:Kmu2}) we need to identify
first $g$ and $\partial h/\partial \dot{x}^\mu$. Then we proceed
to integrate the partial derivatives to obtain $h$, 
and enter this information in (\ref{eq:V2}) to obtain $f$ and, 
finally, we finish the process by inserting these quantities in (\ref{eq:lag4}).

We also note from Eqs.~(\ref{eq:P11}) and (\ref{eq:p11}) that we have
the relation
\begin{equation}
\label{eq:condition-2}
\frac{ \partial P_\nu}{\partial x^\mu} = \frac{\partial \mathfrak{p}_{\mu}}{\partial 
\dot{x}^\nu}. 
\end{equation}
This identity is the key point when we try to connect the two points of view for 
the Lagrangians~(\ref{eq:lag1}) and~(\ref{eq:lag4}). 
From Eqs.~(\ref{eq:Mmunu}), (\ref{eq:pp1}) and~(\ref{eq:condition-2})
we have
\begin{equation}
 M_{\mu \nu} = - \frac{\partial \mathbf{p}_{\mu}}{\partial \dot{x}^\nu}
= - \frac{\partial^2 L_d}{\partial \dot{x}^\mu \partial \dot{x}^\nu}
=M_{\nu \mu}.
\label{eq:Mmunu-2}
\end{equation}
We have thus established that the mass-like matrix (\ref{eq:Mmunu}) corresponds 
to the Hessian matrix for the $s$-equivalent Lagrangian $L_d$. By virtue of the 
symmetry of $M_{\mu \nu}$ we have that
\begin{eqnarray}
\frac{\partial \mathbf{p}_{\mu}}{\partial \dot{x}^\nu} &=&  \frac{\partial 
\mathbf{p}_{\nu}}{\partial \dot{x}^\mu},
\\
\frac{\partial M_{\mu \nu}}{\partial \dot{x}^\rho} &=& \frac{\partial 
M_{\rho \mu}}{\partial \dot{x}^\nu} = \frac{\partial M_{\nu \rho}}{\partial \dot{x}^\mu}.
\label{eq:M-1}
\end{eqnarray}
Moreover, the force vector becomes 
\begin{equation}
F_\mu = \frac{\partial \mathbf{p}_\mu}{\partial x^\nu} \dot{x}^\nu 
+ \frac{\partial^2 g}{\partial x^\mu \partial x^\nu}\,h\,\dot{x}^\nu
- \frac{\partial V}{\partial x^\mu},
\end{equation}
and, by considering (\ref{eq:condition-2}) we have in addition the 
identity
\begin{eqnarray}
\Theta_{\mu \nu}
:= \frac{\partial P_\mu}{\partial x^\nu} -
\frac{\partial P_\nu}{\partial x^\mu}   
= 
\frac{\partial \mathfrak{p}_\nu}{\partial \dot{x}^\mu} - 
\frac{\partial \mathfrak{p}_\mu}{\partial \dot{x}^\nu},
\label{eq:id-3}
\end{eqnarray}
where the last relation establishes that the curl of $P_\mu$ in the coordinate configuration space
sector corresponds to the curl of the non-dynamical momenta $\mathfrak{p}_\mu$
in the velocity configuration space sector.  
%

\section{Ostrogradski-Hamiltonian approach}
\label{sec:ostro}

Since the Lagrangian~(\ref{eq:lag1}) depends on the accelerations, 
it is quite natural that we try to develop a canonical analysis by means of the 
OH approach~\cite{ostro,nesterenko1}. The configuration 
space is spanned by ${\cal C}= \left\lbrace x^\mu; \dot{x}^\mu \right\rbrace $ 
and in consequence the ordinary phase space is enlarged and spanned by $
\Gamma= \left\lbrace x^\mu, p_\mu ; \dot{x}^\mu,P_\mu\right\rbrace$, where 
the conjugate momenta to $\dot{x}^\mu$
and $x^\mu$ are given by~(\ref{eq:P1}) and (\ref{eq:p1}), respectively. 
The highest momenta $P_\mu$ must satisfy the $N$ primary constraints
\begin{equation}
C_\mu = P_\mu - K_\mu (x^\nu,\dot{x}^\nu) \approx 0 ,
\label{eq:C1}
\end{equation}
which follows directly from the definition (\ref{eq:P1}). 
Hereinafter and following Dirac's terminology, the weak equality symbol $\approx$ 
stands for equality modulo all constraints~\cite{Dirac,gauge,rothe}. 
Squaring the primary constraints in order to obtain a resulting 
Hamiltonian quadratic in the 
momenta $P_\mu$ does not prove to be always correct~\cite{nesterenko2}.
The canonical Hamiltonian is 
\begin{equation}
H_0 = p_\mu \dot{x}^\mu + P_\mu \ddot{x}^\mu - L (x^\mu,\dot{x}^\mu,
\ddot{x}^\mu) = p_\mu \dot{x}^\mu - V(x^\mu,\dot{x}^\mu) ,
\label{eq:H0} 
\end{equation}
so that the total Hamiltonian of the system is given by
\begin{equation}
H  = p_\mu \dot{x}^\mu - V + u^\mu\,C_\mu .
\label{eq:Ht}
\end{equation}
Here, $u^\mu$ are Lagrange multipliers enforcing the primary constraints
(\ref{eq:C1}). These are a priori functions of the phase space variables
and possibly of time.

For two phase space functions $F$ and $G$, we introduce the generalized
Poisson bracket (PB)
\begin{equation}
\left\lbrace F,G \right\rbrace := \frac{\partial F}{\partial x^\mu} 
\frac{\partial G}{\partial p_\mu} + \frac{\partial F}{\partial 
\dot{x}^\mu} \frac{\partial G}{\partial P_\mu} - ( F \longleftrightarrow G).
\label{eq:PB} 
\end{equation}
Thus, we observe easily that the primary constraints are in 
involution in a strong sense
\begin{equation}
 \left\lbrace C_\mu ,C_\nu \right\rbrace = 0 .
 \label{eq:PB0}
\end{equation}
From Eqs.~(\ref{eq:C1}) and~(\ref{eq:H0}), it is not hard to find that 
$\{ C_\mu, H_0\} = - \frac{\partial K_\mu}{\partial x^\nu} 
\dot{x}^\nu - p_\mu + \frac{\partial V}{\partial \dot{x}^\mu}
$. Consequently, taking into account~(\ref{eq:Ht}) and~(\ref{eq:PB0}), persistence 
of the primary constraints under evolution of the parameter $\tau$ 
requires that
\begin{equation}
\{ C_\mu, H\} \approx \{ C_\mu, H_0 \} + u^\nu \{ C_\mu, C_\nu\}
= - \frac{\partial K_\mu}{\partial x^\nu} 
\dot{x}^\nu - p_\mu + \frac{\partial V}{\partial \dot{x}^\mu}
\approx 0.
\end{equation}
As a consequence we can identify the $N$ secondary constraints
\begin{equation}
{\cal C}_\mu := p_\mu - \frac{\partial V}{\partial \dot{x}^\mu}
+ \frac{\partial K_\mu }{\partial x^\nu }\,\dot{x}^\nu \approx 0.
\label{eq:C2}
\end{equation}
 
Now, a straightforward calculation shows that the PB
of primary and secondary constraints is directly related to the 
matrix $M_{\mu\nu}$
\begin{equation}
 \left\lbrace {\cal C}_\mu, C_\nu \right\rbrace = M_{\mu \nu}.
\label{eq:PB1} 
\end{equation}
Similarly, the PB of the secondary constraints with the 
canonical Hamiltonian reads $\{ {\cal C}_\mu , H_0\} \\
= -F_\mu$ where $F_\mu$ is given by~(11). Hence, the stationary 
condition on the secondary constraints is given by
\begin{equation}
\{ {\cal C}_\mu, H \} \approx \{ {\cal C}_\mu , H_0 \} + u^\nu \{ {\cal C}_\mu,
C_\nu \} = - F_\mu + u^\nu M_{\mu \nu} \approx 0, 
\label{eq:u}
\end{equation}
which merely provides a restriction on the Lagrange multipliers $u^\mu$, and 
therefore the process of generation of more constraints is finished at this point.
Additionally, the PB among the secondary constraints
\begin{equation}
\left\lbrace {\cal C}_\mu ,{\cal C}_\nu \right\rbrace = \frac{\partial 
p_\nu}{\partial x^\mu} -  \frac{\partial p_\mu}{\partial x^\nu} =: X_{\mu \nu}.
\label{eq:Xmunu}
\end{equation}
This expression stands for the curl of the momenta $p_\mu$ in the coordinate space
sector and it will play a fundamental role in the subsequent discussion. Further, 
$X_{\mu \nu}$ is also invariant under the transformations~(\ref{eq:Ktransf}) 
and (\ref{eq:Vtrans}). In terms of the momenta associated to the 
dynamic Lagrangian (\ref{eq:lag-d}) we have that
\begin{equation}
X_{\mu \nu} = \frac{\partial 
\mathbf{p}_\nu}{\partial x^\mu} - \frac{\partial \mathbf{p}_\mu}{\partial x^\nu}.
\label{eq:Xmunu0}
\end{equation}
As can easily be verified, the matrix (\ref{eq:Xmunu}) satisfies the Bianchi
identity
\begin{equation}
\frac{\partial X_{\mu \nu}}{\partial x^\rho} + 
\frac{\partial X_{\rho \mu}}{\partial x^\nu} + 
\frac{\partial X_{\nu \rho}}{\partial x^\mu}= 0  \,. 
\label{eq:X-1}
\end{equation}

There are other relationships, known as Helmholtz conditions, which 
relate matrices $M_{\mu \nu}$ and $X_{\mu \nu}$ with all the momenta. 
See \ref{appendix} for more details.

\subsection{Hamiltonian equations}

The Hamiltonian equations are obtained in the standard way. First,
\begin{equation}
\ddot{x}^\mu \approx \left\lbrace \dot{x}^\mu, H \right\rbrace
= u^\nu\,\frac{\partial C_\nu}{\partial P_\mu} = u^\mu,
\label{eq:ham1}
\end{equation}
i.e., $u^\mu$ are nothing but the accelerations which were not possible to
solve out from~(\ref{eq:P1}). Notice that Eq.~(\ref{eq:u}) is reduced to 
the eom~(\ref{eq:eom1}) when $u^\mu$ is inserted. The second 
Hamilton equation for $\dot{x}^\mu$ is a mere identity
\begin{equation}
\dot{x}^\mu \approx \left\lbrace {x}^\mu, H \right\rbrace
= \frac{\partial H_0}{\partial p_\mu} = \dot{x}^\mu,
\label{eq:ham2}
\end{equation}
since the only dependence on $p_\mu$ in $H_0$ is through the term $p_\mu \dot{x}^\mu$. 
With respect to the Hamilton equation for $\dot{P}_\mu$ we have
\begin{eqnarray}
\dot{P}_\mu \approx \left\lbrace P_\mu, H \right\rbrace 
=
- \frac{\partial H_0}{\partial \dot{x}^\mu} - u^\nu \, \frac{\partial
C_\nu}{\partial \dot{x}^\mu}
= 
- p_\mu + \frac{\partial V}{\partial \dot{x}^\mu} + u^\nu \frac{\partial K_\nu}
{\partial \dot{x}^\mu}.
\label{eq:ham3}
\end{eqnarray}
This expression reproduces the form of the momenta~(\ref{eq:p1}) once we 
insert the Lagrange multiplier~(\ref{eq:ham1}). Finally, the  
Hamilton equation for $\dot{p}_\mu$ reads
\begin{eqnarray}
\dot{p}_\mu 
\approx 
\left\lbrace p_\mu, H \right\rbrace 
=
- \frac{\partial H_0}{\partial {x}^\mu} - u^\nu \, \frac{\partial
C_\nu}{\partial {x}^\mu}
=  
\frac{\partial V}{\partial {x}^\mu} + u^\nu \frac{\partial K_\nu}
{\partial {x}^\mu}.
\label{eq:ham4}
\end{eqnarray}
Here again, the eom~(\ref{eq:eom2}) are recovered 
when we insert the Lagrange multiplier~(\ref{eq:ham1}). Thus, 
the Lagrangian and Hamiltonian descriptions 
are entirely equivalent, as expected.

\section{Analysis of the constraints}

Following the ordinary Dirac treatment for constrained systems
we require to decompose the set of primary and secondary 
constraints into first- and second-class 
constraints~\cite{Dirac,gauge,rothe}. 
According to Eqs.~(\ref{eq:PB0}), (\ref{eq:PB1}) and (\ref{eq:Xmunu})
we construct the matrix $\Omega_{ij}$ whose elements are the 
PB among all the constraints
\begin{equation}
(\Omega_{ij})=\left(
\begin{array}{cc}
\left\lbrace C_\mu , C_\nu \right\rbrace  &  \left\lbrace C_\mu , 
{\cal C}_\nu \right\rbrace \\
 \left\lbrace {\cal C}_\mu , {C}_\nu \right\rbrace &
\left\lbrace {\cal C}_\mu , {\cal C}_\nu  \right\rbrace
\end{array}
\right) = \left(
\begin{array}{cc}
0_{N \times N} & -M_{\mu \nu} \\
M_{\mu \nu} & X_{\mu \nu}
\end{array}
\right),
\label{eq:Omega}
\end{equation}
where $M_{\mu\nu}$ and $X_{\mu\nu}$ were defined 
in~(\ref{eq:Mmunu}) and~(\ref{eq:Xmunu}), respectively, 
and $i,j=1,2,\ldots,N, N +1, \ldots, 2N$. The rank of this non-singular  
matrix is $2N$ which indicates the presence of $2N$ second-class 
constraints. Indeed, the primary and secondary constraints exhaust 
the whole set of constraints being  second-class and we can use them as 
mere identities strongly equal to zero. This is useful in order to
express some canonical variables in terms of others. The counting of 
degrees of freedom (dof) is readily obtained~\cite{gauge}: 
$\mathrm{dof} = [4N - 2N]/2= N$.  

We may therefore construct the Dirac bracket as usual~\cite{gauge,rothe}. For two 
phase space functions, $F$ and $G$, we define
\begin{equation}
\left\lbrace F, G \right\rbrace^{*} := \left\lbrace F, G \right\rbrace
- \left\lbrace  F, \chi_i \right\rbrace  \Omega^{-1} _{ij}
\left\lbrace \chi_j, G \right\rbrace,
\label{eq:DB} 
\end{equation}
where $\chi_i$ denotes the second-class constraints $\chi_i =
(C_\mu, {\cal C}_\mu)$ and $\Omega^{-1} _{ij}$ is the inverse of the
matrix $\Omega_{ij}$ such that $\Omega^{-1} _{ij}\Omega^j{}_k = 
\delta_{ik}$. As a result we have a reduced phase-space Hamiltonian 
description where we must replace the PB~(\ref{eq:PB}) 
by the Dirac bracket~(\ref{eq:DB}) in order that 
the second-class constraints hold strongly.

It is worthwhile commenting on these results in connection with the inverse
problem of Lagrangian mechanics. The representation (\ref{eq:Omega})
for the non-singular matrix $\Omega_{ij}$ is consistent with the
results found in \cite{sarlet,henneaux2} where $\Omega_{ij}$ takes
the form of a non-singular matrix which makes the eom~(\ref{eq:eom2}) 
derivable from a variational principle. Additionally, 
the components of the matrix $\Omega_{ij}$ follow the Helmholtz integrability
conditions given by equations~(\ref{eq:M-1}), (\ref{eq:X-1}) and~(\ref{eq:h-1}), 
respectively.

\subsection{The singular case}

A wide range of interesting physical systems does not have a regular matrix
$M_{\mu \nu}$. For a singular matrix $M_{\mu \nu}$ of rank $R_M$ there 
exist $n=N-R_M$ independent left (or right) zero-modes eigenvectors 
$\xi^\mu _{(n)}=\xi^\mu _{(n)}(x^\nu, \dot{x}^\nu)$, satisfying
\begin{equation}
 \xi^\mu _{(n)} M_{\mu \nu} = 0 \quad
\mbox{or} \quad  M_{\mu \nu} \xi^\nu _{(n)}= 0.
\label{eq:zero-modes}
\end{equation}
This condition clearly imposes certain restrictions on the previously 
introduced arbitrary function  
$V(x^\mu,\dot{x}^\mu)$ and the momenta $p_\mu$.  Indeed, from eom 
(\ref{eq:eom2}) we have the Lagrangian constraints 
\begin{equation}
\varphi_{(n)}(x^\mu,\dot{x}^\mu) :=  \xi^\mu _{(n)} F_\mu =  \xi^\mu _{(n)} 
\left( \frac{\partial p_\mu}{\partial {x}^\nu}\,\dot{x}^\nu - 
\frac{\partial V}{\partial x^\mu}\right) =0,
\label{eq:varphi}
\end{equation}
for each independent left zero-mode  $\xi^\mu _{(n)}$. On the contrary, 
if these conditions are not satisfied then the action~(\ref{eq:action}) does 
not provide a consistent theory. Therefore, for a singular matrix $M_{\mu\nu}$ 
and a consistent action~(\ref{eq:action}), this description requires that the 
accelerations $\ddot{x}^\mu$ or the Lagrange multipliers $u^\mu$, seen as 
solutions to~(\ref{eq:eom2}) or~(\ref{eq:u}), respectively, are defined only 
up to arbitrary linear combinations of the zero-modes eigenvectors $\xi^\mu _{(n)}$, 
that is, $\ddot{x}^\mu \to \ddot{x}^\mu +  a_n\, \xi^\mu _{(n)}$ and $u^\mu 
\to u^\mu + b_n\,\xi^\mu _{(n)}$ where $a_n$ and $b_n$ are some arbitrary 
constants.

\section{Examples}

We turn now to consider some examples of applications of the approach developed above.

\subsection{Chiral oscillator}
\label{sec:chiral}

Consider the 2-dimensional non-relativistic oscillator with a Chern-Simons 
like term described by the Lagrangian~\cite{lukierski,Gonera1,Gonera2,Acatrinei,Horvathy}
\begin{equation}
L(x^\mu,\dot{x}^\mu,\ddot{x}^\mu) = -\frac{\lambda}{2}\epsilon_{\mu \nu}
\dot{x}^\mu \ddot{x}^\nu + \frac{m}{2} \dot{x}_\mu \dot{x}^\mu,
\label{eq:chiral-L}
\end{equation}
where $\lambda$ and $m$ are non-vanishing constants, $x^\mu = (x,y)$ and 
$\epsilon_{\mu \nu}$ is the Levi-Civita symbol such that $\epsilon_{12} 
= 1$ ($\mu,\nu =1,2$). 
We immediately identify
\begin{eqnarray}
K_\mu &=&  \frac{\lambda}{2}\epsilon_{\mu \nu}
\dot{x}^\nu = \frac{\lambda}{2}(\dot{y},-\dot{x}) ,
\label{eq:Kmu-2}
\\
V &=& \frac{m}{2} \dot{x}_\mu \dot{x}^\mu .
\end{eqnarray}
Notice that both $K_\mu$ and $V$ only depend on the velocities. 
Condition~(\ref{eq:condition}) is satisfied by (\ref{eq:Kmu-2}) and, in 
consequence it is not possible to identify a surface term.  It is crucial to note that 
this is an effect of the fact that condition~(\ref{eq:id-1}) is not 
followed by this system, that is, 
$\partial K_\mu / \partial \dot{x}^\nu \neq
\partial K_\nu / \partial \dot{x}^\mu$, 
which is a signal that eom for this case will not be of second-order.
In fact, the form of the eom~(\ref{eq:eom2}) is not valid anymore for this model. 
Indeed, from (\ref{eq:eom1}) we obtain
\begin{equation}
\lambda \epsilon_{\mu \nu} \dddot{x}^\nu - m \ddot{x}_\mu = 0. 
\end{equation}
Despite the third-order of these equations, it is possible to reduce this
system of equations of motion to a soluble second-order one, in a standard manner. 
From (\ref{eq:P1}) and (\ref{eq:p1}) the OH momenta associated 
to~(\ref{eq:chiral-L}) are
\begin{eqnarray}
P_\mu &=& \frac{\lambda}{2} \epsilon_{\mu \nu} \dot{x}^\nu ,
\\
p_\mu &=& m \dot{x}^\mu - \lambda \epsilon_{\mu \nu} \ddot{x}^\nu .
\end{eqnarray}
The canonical Hamiltonian reads
\begin{equation}
H_0 = p_\mu \dot{x}^\mu - \frac{m}{2} \dot{x}_\mu \dot{x}^\mu.
\end{equation}
The two primary constraints are given by $C_\mu = P_\mu - \frac{\lambda}{2} 
\epsilon_{\mu \nu} \dot{x}^\nu \approx 0$
and, in consequence the total Hamiltonian is
\begin{equation}
H = p_\mu \dot{x}^\mu - \frac{m}{2} \dot{x}_\mu \dot{x}^\mu
+ u^\mu \left( P_\mu - \frac{\lambda}{2} \epsilon_{\mu \nu} \dot{x}^\mu \right),
\end{equation}
where $u^\mu$ are Lagrange multipliers enforcing the primary constraints.
The PB among the primary constraints results
\begin{equation}
\left\lbrace C_\mu, C_\nu \right\rbrace = \lambda\,\epsilon_{\nu \mu}. 
\end{equation}
Evolution in time of $C_\mu$ leads only to a restriction of $u^\mu$
\begin{equation}
\dot{C}_\mu = \left\lbrace C_\mu , H \right\rbrace
\approx - p_\mu + m \dot{x}_\mu + \lambda \, \epsilon_{\nu \mu} u^\nu
= 0,
\end{equation}
and in consequence $C_\mu$ are second-class constraints. The number of dof is 
readily obtained: $\mathrm{dof}= [8 - 2]/2 = 3$.
See \cite{lukierski} for a detailed quantum description for this model.

\subsection{Geometric dynamics inferred from a second-order form}

Consider the so-called {\it second-order general energy Lagrangian} given 
by \cite{Foster,udriste}
\begin{equation}
L (x^\mu, \dot{x}^\mu , \ddot{x}^\mu )= \omega_\mu (x^\alpha) \ddot{x}^\mu
+ \omega_{\mu \nu} (x^\alpha) \dot{x}^\mu \dot{x}^\nu, 
\end{equation}
where the terms $\omega_\mu$ are considered as potentials on a Riemannian manifold
$(\mathbb{R}^n, \delta_{\mu \nu})$ 
with the metric $\delta_{\mu \nu}$, 
and the terms $\omega_{\mu\nu}$
are symmetric functions on the indices.
These coefficients $(\omega_\mu,\omega_{\mu \nu})$ characterize the 
second-order form $\omega = \omega_\mu (x^\alpha) d^2 x^\mu + \omega_{\mu \nu}(x^\alpha)
dx^\mu \otimes dx^\nu$.   The geometric properties of this Lagrangian have been 
extensively studied in a huge variety of contexts ranging from biomathematics,
mathematical economics, to a non-standard version of electrodynamics~\cite{Foster,udriste,2ndorder}.  
In fact, this Lagrangian is interesting as the inherent eom are straightforwardly 
related to the geodesic equation. The tensor 
\begin{equation}
N_{\mu \nu} := \frac{1}{2} (\partial_\nu \omega_\mu + \partial_\mu 
\omega_\nu  ),
\end{equation}
which is known as the {\it deformation rate tensor field}, allows us to
introduce a new metric
\begin{equation}
g_{\mu \nu} := \omega_{\mu \nu} - N_{\mu \nu},
\end{equation}
where we assume that $\mbox{det} (g_{\mu \nu}) \neq 0$. 
In our notation, $K_\mu = \omega_\mu$ and $V = \omega_{\mu \nu} 
\dot{x}^\mu \dot{x}^\nu$. Note further that (\ref{eq:id-1}) is
trivially satisfied. From (\ref{eq:P1}) and (\ref{eq:p1}) we have
\begin{eqnarray}
P_\mu &=& \omega_\mu , 
\\
p_\mu &=& (2 \omega_{\mu \nu} - \partial_\nu \omega_\mu)\dot{x}^\nu =
(2 g_{\mu \nu} + \partial_\mu \omega_\nu)\dot{x}^\nu .
\end{eqnarray}
The mass-like matrix (\ref{eq:Mmunu}) becomes $M_{\mu \nu} = -2 g_{\mu \nu}$. Similarly,
the force vector is
\begin{equation}
F_\mu = 2 (\omega_{\alpha \beta \mu} - \Omega_{\alpha \beta \mu})
\dot{x}^\alpha \dot{x}^\beta,
\end{equation}
where we have introduced the connection symbols associated to 
$\omega_{\mu \nu}$
\begin{eqnarray}
\omega_{\alpha \beta \mu} &:=& \frac{1}{2} (\partial_\alpha \omega_{\beta \mu}
+ \partial_\beta \omega_{\alpha \mu} - \partial_\mu \omega_{\alpha \beta}) ,
\\
\Omega_{\mu \alpha \beta} &:=& \frac{1}{2} ( \partial_\alpha \partial_\beta 
\omega_\mu + \partial_\mu \partial_\beta \omega_\alpha - \partial_\mu 
\partial_\alpha \omega_\beta ). 
\end{eqnarray}
Thus, from (\ref{eq:eom2}), the equations of motion read
\begin{equation}
g_{\mu \alpha} \ddot{x}^\alpha + \Gamma_{\alpha \beta \mu} \dot{x}^\alpha 
\dot{x}^\beta = 0,
\end{equation}
where $\Gamma_{\alpha \beta \mu}:= \omega_{\alpha \beta \mu} - \Omega_{\alpha
\beta \mu}$. Technically, as discussed in~\cite{udriste}, we have 
obtained the dynamics of the geodesics governed by an
Otsuki type connection $\Gamma_{\alpha \beta \mu}$~\cite{otsuki}. Further, 
from relations~(\ref{eq:lag-d}) 
and~(\ref{eq:lag-s}), we realize that 
the dynamic and the surface 
Lagrangians are explicitly given 
by $L_d=g_{\mu\nu}\dot{x}^\mu\dot{x}^\nu$
and $L_s=\omega_\mu \ddot{x}^\mu +
N_{\mu\nu}\dot{x}^\mu\dot{x}^\nu$,
respectively. The canonical Hamiltonian is
\begin{equation}
 H_0 = p_\mu \dot{x}^\mu - \omega_{\mu \nu} \dot{x}^\mu \dot{x}^\nu,
\end{equation}
and we have $N$ primary constraints
\begin{equation}
 C_\mu = P_\mu - \omega_\mu \approx 0,
 \label{eq:pC}
\end{equation}
which are in involution in a strong sense, $\left\lbrace  
C_\mu , C_\nu \right\rbrace= 0$. The total Hamiltonian is $H=H_0 +
u^\mu C_\mu$. By requiring that 
$\dot{C}_\mu\approx 0$, we generate the $N$ secondary constraints
\begin{equation}
 {\cal C}_\mu = p_\mu - ( 2 \omega_{\mu \nu} - \partial_\nu \omega_\mu ) 
\dot{x}^\nu \approx 0.
\label{eq:sC}
\end{equation}
The PBs among the constraints~(\ref{eq:pC}) and (\ref{eq:sC}) read
$\left\lbrace C_\mu , {\cal C}_\nu \right\rbrace = - 2g_{\mu \nu}$.
As discussed before, no further constraints exist as evolution in time of 
the secondary constraints ${\cal C}_\mu\approx 0$ leads
to an expression that determines the Lagrange multipliers
\begin{equation}
2 g_{\mu \nu} (\ddot{x}^\mu - u^\mu) \approx 0. 
\end{equation}
Indeed, $u^\mu = \ddot{x}^\mu$, in agreement with~(\ref{eq:ham1}).
In addition, we have $\left\lbrace {\cal C}_\mu , {\cal C}_\nu \right\rbrace = X_{\mu \nu}
= 2 (\partial_\mu g_{\nu \rho} - \partial_\nu g_{\mu \rho})\dot{x}^\rho$.
Hence, from Eq.~(\ref{eq:Omega}) we have
\begin{equation}
(\Omega_{ij})= \left(
\begin{array}{cc}
0 & 2g_{\mu \nu} \\
-2g_{\mu \nu} & 2 (\partial_\mu g_{\nu \rho} - \partial_\nu g_{\mu \rho})\dot{x}^\rho
\end{array}
\right) \,,
\label{eq:Omega2}
\end{equation}
and its inverse matrix
\begin{equation}
(\Omega_{ij} ^{-1})= \frac{1}{4} \left(
\begin{array}{cc}
X_{\alpha \beta}g^{\alpha \mu} g^{\beta \nu} & -2g^{\mu \nu} \\
2g^{\mu \nu} & 0
\end{array}
\right),
\label{eq:Omega3}
\end{equation}
where $g^{\mu\nu}$ stands for the inverse matrix
associated to the metric $g_{\mu\nu}$.
It follows that the corresponding Dirac brackets become
\begin{eqnarray}
\left\lbrace   F, G \right\rbrace^* 
&=&  
\left\lbrace   F, G \right\rbrace
- \frac{1}{4} X_{\alpha \beta}g^{\alpha \mu} g^{\beta \nu} \left\lbrace   F, 
C_\mu \right\rbrace \left\lbrace C_\nu, G \right\rbrace
\nonumber
\\
& & 
+\frac{1}{2} g^{\mu \nu} \left( \left\lbrace   F, 
C_\mu \right\rbrace \left\lbrace {\cal C}_\nu, G \right\rbrace -
\left\lbrace   F, {\cal C}_\mu \right\rbrace \left\lbrace C_\nu, G \right\rbrace \right) . 
\end{eqnarray}

\subsection{Optimal growth model in Economics}

We start by considering the optimal growth model as considered in~\cite{udriste,Tu}.
The model consists of maximizing the functional $J[x]:= \int_{\mathbb{R}} U(x,\dot{x},\ddot{x})dt$,
where the utility function $U$ will work as our Lagrangian through the combination
$U(C,\dot{C}) = C^a + \gamma \dot{C}$, and is considered as an increasing and concave
function. Here the variable $x$ stands for the capital stock, while $C(x,\dot{x})$ stands
for the consumption function regarding the dynamic utility and capital accumulation,
while the parameters $a,\gamma \in [0,1]$. Following~\cite{udriste}, we consider
that $C(x,\dot{x})$ follows the capital accumulation law $C(x,\dot{x}) = Y(x) - \dot{x}$
where $Y(x)$ stands for the Gross National Income. According to the Economics lore,
$C(x, \dot{x})$ is the Gross National Product left over after the accumulation
$\dot{x}$ is met. Under the assumption that $Y(x) = b x$ with $b$ being a constant, the 
utility consumption function $U(C,\dot{C})$ may be rewritten as
\begin{equation}
U(x,\dot{x}, \ddot{x}) = (bx - \dot{x})^a + \gamma b \dot{x} - 
\gamma \ddot{x},
\label{eq:U}
\end{equation}
which clearly defines an economic growth model affine in the acceleration term 
$\ddot{x}$. By comparison to (\ref{eq:lag1}), we immediately identify
\beq
K &=& -\gamma,
\\
V&=& (bx - \dot{x})^a + \gamma b \dot{x}.
\eeq
From Eqs.~(\ref{eq:lag-d}) and (\ref{eq:lag-s}) we may recognize the dynamical and surface
parts of the Lagrangian $U$ as $U_d = V(x,\dot{x})$ and as $U_s = -\gamma \ddot{x}$,
respectively. Also, from (\ref{eq:eom2}) the eom reads $(1-a)\ddot{x}
+ (a-2)b\dot{x} + b^2 x = 0$. The momenta are
\beq
P&=& -\gamma,
\\
p&=& - a(bx - \dot{x})^{a-1} + \gamma b,
\eeq
while the canonical Hamiltonian is
\begin{equation}
H_0 = (p - \gamma b)\dot{x} - (bx - \dot{x})^a. 
\end{equation}
From Eqs.~(\ref{eq:C1}) and (\ref{eq:C2}) we have the primary constraint $C = P+\gamma \approx 0$
and the secondary constraint ${\cal C} = p + a(bx - \dot{x})^{a-1} - \gamma b \approx 0$
which results second-class because $\{ C, {\cal C}\} = a (a-1) (bx - \dot{x} )^{a-2}$. Thus, the 
total Hamiltonian governing the evolution in the reduced phase space is $H_0$ itself.
Additionally, from the second-class ${\cal C}$ we can obtain $\dot{x}$ in terms of $x$ and 
$p$ such that the total Hamiltonian in the reduced phase space reads
\begin{equation}
H = (p-\gamma b)bx - (1-a) \left( \frac{a}{\gamma b} \right)^{\frac{a}{1-a}} \left( 1
- \frac{p}{\gamma b} \right)^{\frac{a}{a-1}}.
\label{eq:H-economics-1}
\end{equation}
The number of dof is: dof $= [4-2]/2=1$, which
corresponds with the capital stock $x$.

An interesting approximation is provided whenever the capital accumulation results lower 
than the capital stock. For such a case we have $|p/\gamma b| <1$ and thus, by expanding
the second term in (\ref{eq:H-economics-1}) up to quadratic terms of the quotient $p/\gamma b$,
the approximated total Hamiltonian may be written as the expression
\begin{equation}
 H = \frac{1}{2m} \left[ p + m A(x) \right]^2 + V(x),
\label{eq:H-economics-2}
 \end{equation}
where we have made the identifications
\beq
m &:=& \left[ \frac{1}{a(a-1)} \left( \frac{a}{\gamma b} \right)^{\frac{2-a}{1-a}} \right]^{-1},
\\
A(x) &:=& bx - \left( \frac{a}{\gamma b} \right)^{\frac{1}{1-a}},
\eeq
and the potential function $V(x)$ is recognized as the quadratic function
\begin{equation}
V(x) = - \left[ \gamma b^2 x + (1-a) \left( \frac{a}{\gamma b} \right)^{\frac{a}{1-a}} \right] 
- \frac{m}{2} A(x)^2.
\end{equation}
Hamiltonian~(\ref{eq:H-economics-2}) resembles the one usually encountered for a charged
particle with a mass-like term $m$ in a magnetic field provided by $A(x)$ and under a
quadratic interaction given by the potential $V(x)$. This resemblance may be exploited
in order to straightforward obtain a quantum counterpart for this model.

\section{Covariant theories}
\label{sec:covariant}

Extended objects, sometimes referred to as branes, are 
intended to represent physical systems at almost all energy 
scales~\cite{capo95}. The dynamical evolution is described by an action
constructed with geometrical scalars associated to their surface of 
evolution. The Lagrangian is given by a second-order derivative function, 
$L(g_{ab},K_{ab})$, where $g_{ab}$ and $K_{ab}$ denote the induced metric 
and the extrinsic curvature, respectively,  defined on the surface 
which contains the only significant derivatives of the field variables 
being the immersion functions $X^\mu$~\cite{capo95}. The main symmetry
of this type of actions is the invariance under reparametrizations
of the trajectory of the extended object. Tipically, 
these  geometric models lead to fourth-order equations of motion. Favorably, 
there is a special subset of such geometric invariants whose associated 
equations of motion remain of second-order in $X^\mu$~\cite{ostro-prd,membrane}. 
When these Lagrangians are specialized to a particular geometry they 
clearly exhibit a linear dependence in the accelerations, $\ddot{X}^\mu$.

One of the most important consequences of gauge theories invariant under 
reparametrizations is that the Hamiltonian vanishes identically.
From an uncommon but interesting geometric point of view, the vanishing of the 
Hamiltonian is a consequence of Zermelo conditions~(see~\cite{Kondo,MironL,MironH} 
and~\ref{Noether}) which state the invariance of the Lagrangian under the Lie derivative 
of the Liouville vector fields associated to a second-order Lagrangian~\cite{leon-rds}, that is,
$\pounds_{\Gamma^{(i)}}(L)=0$ for any Liouville vector field $\Gamma^{(i)}$.  In other words,
Zermelo conditions are the necessary conditions for an action integral (in our case~(\ref{eq:action})) 
to be invariant under parametrization of the trajectory $c$ (see~\ref{Noether} for 
further details). On physical grounds, for our case, the action integral (\ref{eq:action}) together with 
(\ref{eq:lag1}), must be a Lorentz invariant as well as a homogeneous function 
in the generalized velocities. Indeed, from Eqs.~(\ref{eq:Gamma1}) and~(\ref{eq:Gamma2})
the associated Legendre transformation yields
\begin{eqnarray}
H_0 &=&
\frac{\partial L}{\partial \ddot{x}^\mu} \ddot{x}^\mu +
\left[ \frac{\partial L}{\partial \dot{x}^\mu} - \frac{d}{d\tau}\left(
\frac{\partial L}{\partial \ddot{x}^\mu} \right)  \right]\dot{x}^\mu 
 - L,
\nn 
\\
&=& 
\Gamma^{(2)} (L) - L - \frac{d}{d\tau} \Gamma^{(1)}(L),
\nn
\end{eqnarray}
or, more explicitly
\begin{equation}
H_0 =
\frac{\partial V}{\partial \dot{x}^\mu} \dot{x}^\mu -
\frac{\partial K_\mu}{\partial x^\nu} \dot{x}^\nu \dot{x}^\mu - V
= {\cal E}_c ^{(2)}.
\label{eq:H00}
\end{equation}
Note that the Hamiltonian $H_0$ is equivalent to the energy $\mathcal{E}^{(2)}_c$
introduced in relation~(\ref{eq:energy2}), which, together with the energy 
$\mathcal{E}_c^{(1)}$, vanishes whenever Zermelo conditions are considered.
Zermelo conditions~(\ref{eq:Zermelo}) may be interpreted in our notation 
as the condition that the function $V$ has to be a homogeneous function 
of first degree in the velocities, while the function $K_\mu$
has to be orthogonal to the velocity vector.  Under these conditions,  
our canonical Hamiltonian vanishes accordingly as shown in~(\ref{eq:GenEnergy2}).
With regards the Noether charge~(\ref{eq:Q-noether}), this reduces immediately to
\begin{equation}
{\cal Q} (L, \phi) = p_\mu \dot{x} + P_\mu \ddot{x}^\mu - \phi. 
\end{equation}

A different but equivalent viewpoint shows also the vanishing of the canonical 
Hamiltonian $H_0$ in terms of the functions $f$ and $h$ introduced in~(\ref{eq:lag4}). Indeed, from 
Eq.~(\ref{eq:lag-d}) we see that the function $f(x^\mu,\dot{x}^\mu)$ must be also homogeneous 
of first degree in the velocities, while it is required that the function $h(\dot{x}^\mu)$ 
be homogeneous of order zero, that is, 
\begin{equation}
\dot{x}^\mu \frac{\partial h }{\partial \dot{x}^\mu} = 0.
\label{eq:identity-normal}
\end{equation}
By virtue of $\dot{x}^\mu \partial f / \partial \dot{x}^\mu = f$, we have 
that $\dot{x}^\nu (\partial^2 f/\partial \dot{x}^\mu \partial \dot{x}^\nu) 
= - \dot{x}^\nu M_{\mu \nu} = 0$ and in consequence the velocity vector 
results to be one of the zero-modes of $M_{\mu \nu}$. This fact is important 
since if $M_{\mu \nu}$ is singular and if the rows (or columns) are proportional to
each other, hence $M_{\mu \nu}$ can be written as the direct product of 
two vectors. For our case,
\begin{equation}
 M_{\mu \nu} = l(x^\alpha,\dot{x}^\alpha)\,n_\mu n_\nu,
\label{eq:matrixM}
\end{equation}
where $n_\mu$ is a unit spacelike vector such that $n_\mu \dot{x}^\mu=0$
and $n_\mu n^\mu =1$, and $l(x^\alpha,\dot{x}^\alpha)$ is a
function of the configuration space.

In the light of this geometrical interpretation, we can see directly from 
Eqs.~(\ref{eq:P11}) and (\ref{eq:pp1}) that contraction of the momenta 
$P_\mu$ and $p_\mu$ with the velocity vector leads to the vanishing of 
$P_\mu \dot{x}^\mu$ as well as to the vanishing of the canonical Hamiltonian 
$H_0$, as given by (\ref{eq:H00}) and (\ref{eq:H0}). This, of course, is a direct
result from Zermelo's conditions~(\ref{eq:Zermelo}).  We also comment 
that these important 
relations are classified as first-class constraints in the Dirac approach 
for constrained systems (see for example, Theorem 1.3 in \cite{gauge}). 
As a byproduct, from Eq.~(\ref{eq:identity-normal}), it follows that $\partial 
h/\partial \dot{x}^\mu$ is proportional to a normal vector, $n^\mu$, satisfying 
also $n_\mu \dot{x}^\mu =0$,
\begin{equation}
\frac{\partial h }{\partial \dot{x}^\mu} = m (\dot{x}^\nu )\, n_\mu, 
\label{eq:h-m}
\end{equation}
where $m$ is function depending on the velocities. This fact is also useful in 
the search of more constraints. All these features represent the hallmark of 
reparametrization invariant systems such as GR and brane 
theories~\cite{ostro-prd,membrane,Dutt,hambranes}.

We should also note the fact that $H_0$ appears as a secondary 
constraint may be inferred initially
from equation~(\ref{eq:varphi}) by considering $\xi^\mu = 
\dot{x}^\mu$. Clearly, 
\begin{eqnarray}
\dot{x}^\mu F_\mu 
= 
\dot{x}^\mu \frac{\partial}{\partial x^\mu}
(p_\nu \dot{x}^\nu - V)
= 
\frac{d}{d\tau}  ( p_\mu \dot{x}^\mu - V ) -
\frac{\partial ( p_\nu \dot{x}^\nu - V ) }{\partial \dot{x}^\mu}\,\ddot{x}^\mu. 
\nonumber
\end{eqnarray}
By using Eqs.~(\ref{eq:Mmunu}), (\ref{eq:P1}) 
and (\ref{eq:p1}) we 
 verify the identity
$\partial(p_\nu \dot{x}^\nu- V)/\partial\dot{x}^\mu =
-M_{\mu\nu}\dot{x}^\nu=0$, which simplifies the last equation 
to
\begin{equation}
\dot{x}^\mu F_\mu = \frac{d}{d\tau} ( p_\mu \dot{x}^\mu - V ).
\label{eq:H0-lag}
\end{equation}
Thus, we have outlined the constrained scheme for the Lagrangian (\ref{eq:lag1})
which will be confirmed shortly by means of Dirac formalism for constrained
systems.

\subsection{Covariant brane theories}
\label{sec:covariant-brane}

In the representation provided by (\ref{eq:Omega}), under certain conditions,
the whole set of second-class constraints $\chi_i$ could involve a hidden sector of 
first-class constraints. Certainly, for covariant theories the matrix 
$M_{\mu \nu}$ is singular and we can uncover these constraints as follows. 
From equation (\ref{eq:zero-modes}) we infer that
\begin{equation}
f^1 _{(n)} := \xi^\mu _{(n)} C_\mu , 
\label{eq:f11}
\end{equation}
are $n$ first-class constraints. Indeed, by contracting equations~(\ref{eq:PB0}) and 
(\ref{eq:PB1}) on the right with $\xi^\mu _{(n)}$, we have $\left\lbrace C_\mu, 
f^1 _{(n)} \right\rbrace \approx 0$ and $\left\lbrace {\cal C}_\mu, f^1 _{(n)} 
\right\rbrace \approx 0$. We note that these identities hold off-shell. Note 
further that still we are left with $R_M$ second-class constraints.

Likewise, we can uncover another set of first-class constraints. To find this, 
we  first project~(\ref{eq:PB1}) on the left by~$\xi^\mu _{(n)}$ and we
obtain that $\left\lbrace f^2 _{(n)} , {C}_\mu \right\rbrace \approx 0$ where 
\begin{equation}
f^2 _{(n)} := \xi^\mu _{(n)} {\cal C}_\mu. 
\end{equation}
Similarly, from~(\ref{eq:zero-modes}),~(\ref{eq:PB1}) and~(\ref{eq:f11})
we have $\left\lbrace f^1 _{(n)} , f^2 _{(n')} \right\rbrace \approx 0$ where
$n,n' = 1,2,\ldots,N-R_M$.

With respect to the second-class constraints, for this specific case and 
adopting a geometric viewpoint, we need to find the whole set of orthogonal 
vectors to $\xi^\mu _{(n)}$, say $n^\mu _{(s)}$, and then contract $C_\mu$ 
and ${\cal C}_\mu$ with $n^\mu _{(s)}$. Here, $s$ keeps track of the number
of orthogonal vectors to $\xi^\mu _{(n)}$. To prove this, we rely on 
a geometrical identity that relates  the complete orthogonal basis 
vectors~\cite{note}. Subsequently, we rewrite the primary and secondary 
constraints by using the metric in order to expand these constraints in terms 
of the orthogonal basis, i.e., $C_\mu = g_\mu{}^\nu C_{\nu} = {\cal H}_\mu {}^\nu 
C_\nu + \perp_\mu{}^\nu C_\nu$. From here, by linear independence, we can identify 
an equivalent set of constraints to $C_\mu$ and ${\cal C}_\mu$. We can check by 
straightforward computation that the resulting expressions $C_\mu n^\mu _{(s)}=0$
and  ${\cal C}_\mu n^\mu _{(s)}=0$ are second-class constraints. Hence, the vectors 
$\xi^\mu _{(n)}$ span a basis for the first-class constraint surfaces in 
the phase space whereas $n^\mu _{(s)}$ span a basis for the second-class
constraint surfaces.

As we have noted for covariant theories, it is convenient to deal with the primary 
and secondary constraints transformed, instead 
of considering them in their original form, by projecting 
them along suitable independent vectors $Z^\mu _{(r)}= Z^\mu _{(r)} (x^\nu,
\dot{x}^\nu)$ in order to have the 
constraints~\cite{nesterenko1}
\begin{eqnarray}
P_\mu Z^\mu _{(r)} &=& A_{(r)} (x^\mu,\dot{x}^\mu),
\\
p_\mu Z^\mu _{(r)} &=& B_{(r)} (x^\mu,\dot{x}^\mu), 
\end{eqnarray}
where $r$ denotes the dimension of the orthonormal basis.
Thus, we have changed to an equivalent set of constraints
\begin{eqnarray}
C_\mu \longrightarrow \phi_\mu &=& A_\mu{}^\nu \,C_\nu,
\label{eq:newc1}
\\
{\cal C}_\mu \longrightarrow \varphi_\mu &=& B_\mu{}^\nu\,
{\cal C}_\nu, 
\label{eq:newc2}
\end{eqnarray}
with $\phi_\mu = (f^1 _{(n)} , \chi^{1} _{(s)})$ and $\varphi_\mu = (f^2 _{(n)}, 
\chi^2 _{(s)})$.
 
\subsection{Electrically charged bubble}
\label{sec:bubble}

As a final illustration we consider a relativistic bubble in the presence 
of an electromagnetic field with a total electric charge $q$ on the 
shell~\cite{membrane,onder}. The Lagrangian is given by
\begin{equation}
L (\dot{t},\ddot{t},r,\dot{r},\ddot{r}) = - \alpha \frac{r^2}{N^2} 
\left(  \ddot{r} \dot{t} -  \dot{r}  \ddot{t} \right) - 2\alpha r\dot{t}
-\beta \frac{q^2 \dot{t}}{r} ,
\label{eq:membrane-L} 
\end{equation}
where $\alpha$ and $\beta$ are constants. Here, $N = \sqrt{\dot{t}^2 - \dot{r}^2}$. 
For this case $\mu, \nu = 1,2= t,r$. We recognize from (\ref{eq:membrane-L})
\begin{eqnarray}
K_1 (\dot{t},r,\dot{r}) &=& K_t = \alpha \frac{r^2 \dot{r}}{N^2},
\\
K_2 (\dot{t},r,\dot{r}) &=& K_r = - \alpha \frac{r^2 \dot{t}}{N^2},
\\
V(\dot{t},r, \dot{r}) &=& - 2\alpha r\dot{t} - \beta \frac{q^2 \dot{t}}{r} .
\end{eqnarray}
From condition~(\ref{eq:condition}) we obtain
that 
$d(\dot{x}^\mu K_\mu)/d\tau = 0$ 
 shows an inconsistency and 
thus it is possible to identify a surface term~\cite{membrane}.
For this case, from Eq.~(\ref{eq:Kmu2}) we observe that $g(r) 
= \alpha r^2$ while the derivatives of the 
function $h$ are $\partial h(\dot{t},\dot{r}) / \partial \dot{t} = \dot{r}/N^2$
and $\partial h(\dot{t},\dot{r}) / \partial \dot{r} = - \dot{t}/N^2$. Thus, 
integrating we have up to a constant, $h(\dot{t},\dot{r}) = - \tanh^{-1} 
(\dot{r}/\dot{t})$. In summary, we have
\begin{equation}
g(r) = \alpha r^2 \quad \mbox{and} \quad \frac{\partial h}{\partial \dot{x}^\mu} =
- \frac{1}{N} n_\mu , 
\end{equation}
where $n_\mu = \frac{1}{N}(-\dot{r}, \dot{t})$. From Eq. (\ref{eq:V2}) we have now 
$f(r,\dot{t},\dot{r}) = 2\alpha r \dot{r} \tanh^{-1}(\dot{r}/\dot{t})
- 2\alpha r \dot{t} - \beta\,q^2\,\dot{t}/r$. Finally, from~(\ref{eq:lag-s}) we obtain 
the associated surface Lagrangian 
\begin{equation}
L_s = \frac{d}{d\tau} \left[ - \alpha r^2 \tanh^{-1}\left(  
\frac{\dot{r}}{\dot{t}}\right)  \right] , 
\end{equation}
which is in agreement with the results found in \cite{membrane}.

From Eqs.~(\ref{eq:P1}) and (\ref{eq:p1}) we have the momenta associated 
to this theory
\begin{eqnarray}
P_1 &=& P_t = \alpha \frac{r^2 \dot{r}}{N^2} ,
\label{eq:Ptelectron}
\\
P_2 &=& P_r = - \alpha \frac{r^2 \dot{t}}{N^2} ,
\label{eq:Prelectron}
\\
p_1 &=& p_t = - \frac{2\alpha r \dot{t}^2}{N^2} - \beta \frac{q^2}{r}
=: -\Omega ,
\label{eq:ptelectron}
\\
p_2 &=& p_r = \frac{2\alpha r \dot{r} \dot{t}}{N^2},
\label{eq:prelectron}
\end{eqnarray}
where $\Omega$ is the conserved bulk energy.
Then, by considering these momenta in (\ref{eq:Mmunu}) and (\ref{eq:Xmunu})
it is found that
\beq
(M_{\mu \nu}) & = & - \frac{4\alpha r \dot{t}}{N^4}  \left(
\begin{array}{cc}
\dot{r}^2 & - \dot{r} \dot{t}\\
- \dot{r} \dot{t}&  \dot{t}^2
\end{array}
\right) \,,\nn\\
(X_{\mu \nu}) & = &  \left(
\begin{array}{cc}
0 & \frac{2\alpha \dot{t}^2}{N^2} - \frac{\beta q^2}{r^2}\\
-  \frac{2\alpha\dot{t}^2}{N^2} + \frac{\beta q^2}{r^2}  & 0
\end{array}
\right) .
\label{eq:matrixM-X}
\eeq
Note that $(M_{\mu \nu})$ is singular. In consequence, we have a left (right) 
zero-mode given by $ \xi^\mu = \dot{x}^\mu = (\dot{t},\dot{r})$. Hence, from (\ref{eq:C1}), 
(\ref{eq:Ptelectron}) and (\ref{eq:Prelectron}) we have a first-class constraint 
\begin{equation}
f_1 = \xi^\mu C_\mu = P_\mu \dot{x}^\mu = P_t \dot{t} + P_r \dot{r} \approx 0 .
\label{eq:f1}
\end{equation}
Similarly, by contracting (\ref{eq:C2}) and considering (\ref{eq:ptelectron}) and 
(\ref{eq:prelectron}) we have another first-class constraint 
\begin{equation}
f_2 = \xi^\mu {\cal C}_\mu = p_\mu \dot{x}^\mu 
+ \left( 2\alpha r + \beta \frac{q^2}{r} \right) \dot{t}= 
p_t \dot{t} + p_r \dot{r} + \left( 2\alpha r + \beta \frac{q^2}{r} \right) 
\dot{t} =0 .
\label{eq:f2}
\end{equation}
An orthogonal vector to $ \xi^\mu = \dot{x}^\mu$, under a Minkowski metric, 
is provided by the unit spacelike vector $n^\mu = \frac{1}{N}(\dot{r}, \dot{t})$. In 
consequence, the matrix $M_{\mu \nu}$ in (\ref{eq:matrixM-X}) may
be expressed as 
\begin{equation}
M_{\mu \nu} = - \frac{4\alpha  r \dot{t}}{N^2}\,n_\mu n_\mu  \,,  
\end{equation}
which is in agreement with Eq.~(\ref{eq:matrixM}).
Hence, by contracting (\ref{eq:C1}) and (\ref{eq:C2}) along $n^\mu$, and 
considering equations (\ref{eq:Ptelectron}) to (\ref{eq:prelectron}), we find the 
second-class constraints
\begin{eqnarray}
s_1 & = & NP_\mu n^\mu + \alpha r^2 = P_t \dot{r} + P_r \dot{t} + \alpha r^2 \approx 0,
\\
s_2 & = & Np_\mu n^\mu + \beta\frac{q^2}{r}= p_t \dot{r} + p_r \dot{t} 
+ \beta \frac{q^2}{r} \approx 0.
\end{eqnarray}
We also note that if we calculate either the PB 
or the Dirac bracket between the first-class constraints,
$f_1$ and $f_2$, we explicitly get $\{f_1,f_2\}=-f_2$
which resembles a truncated Virasoro algebra of the 
form $\{L_m,L_n\}=(m-n)L_{m+n},\ (m=0,n=1)$ by redefining 
the constraints as $L_0 := f_1$ and $L_1 := f_2$. 

The canonical Hamiltonian is
\begin{equation}
H_0 = p_t \dot{t} + p_r \dot{r} + \left( 2\alpha r + \beta 
\frac{q^2}{r}\right)\dot{t},
\end{equation}
and the total Hamiltonian is given by
\begin{equation}
H = H_0 + u^t \left( P_t - \alpha \frac{r^2 \dot{r}}{N^2}\right)
+ u^r \left( P_r + \alpha \frac{r^2 \dot{t}}{N^2}\right).
\end{equation}
where $u^t$ and $u^r$ are Lagrange multipliers enforcing the primary 
constraints~(\ref{eq:Ptelectron}) and~(\ref{eq:Prelectron}), respectively.

\section{Concluding remarks}

We have presented a Hamiltonian analysis for Lagrangians linearly depending 
on the accelerations.  Our presentation was strongly based on the geometric 
analysis of the quantities involved and, in particular, on the conditions 
under which our Lagrangian accepted a decomposition into a true dynamic term 
plus a surface term. In this sense, we analyzed in detail the relation of our 
original setup to the standard Dirac formalism for first-order theories.
Also, we have obtained the general Noether charge for 
this type of systems. 
We have highlighted the role that the surface term played in our formulation, 
and the existence of surface equivalent Lagrangians. The main interest for the 
study of these systems has been motivated by certain brane models, 
although several other examples with different characteristics may be found in 
the literature, for which the existing canonical approaches are not entirely
transparent. In this sense, we emphasize that most of the available 
examples enclose regular Lagrangians, for which the Hamiltonian is developed by 
the introduction of auxiliary variables resulting, from our point of view, in 
a cumbersome description hiding the true geometric interpretation available for 
these sort of systems.  Our claim is that the description developed in this work 
allowed us, in a very natural way, to obtain within the well-known Ostrogradski-Hamilton 
formalism a simplified Hamiltonian version for which the geometric invariants are 
explicitly written. Our geometric formulation allowed us to straightforwardly 
incorporate our results  to the analysis of either non-regular or covariant 
Lagrangians. As stated in the Introduction, we argue that our formulation paves 
the way for the quantisation of this sort of systems, at least within canonical 
schemes of quantization.  We will develop quantum aspects for theories with Lagrangians 
affine in acceleration somewhere else.

\section*{Acknowledgments}
Special thanks to Eloy Ay\'on-Beato and Jasel Berra for useful comments and suggestions.
ER thanks A. P. Balachandran for the encouragement to the paper. ER also acknowledges 
partial support from grant PROMEP, CA-UV: \'Algebra, Geometr\'\i a y Gravitaci\'on. 
MC acknowledges support from a CONACyT scholarship (M\'exico) under the grant 
Repatriaciones, Convocatoria 2015-cuarta fase. AM acknowledges financial support from 
PROMEP UASLP-PTC-402 and from CONACYT-M\'exico under project CB-2014-243433.
This work was partially supported by SNI (M\'exico). 

\appendix
\section{Noether theorem and energies of the system}
\label{Noether}

This section follows closely the notation in 
references~\cite{MironL,MironH,MironNoether}.  
We start by introducing a differentiable vector field $W^\mu$
along the trajectory $c$ at which the action~(\ref{eq:action}) takes place.
We impose that 
$W^\mu$ is (at least locally) regular and 
satisfies the condition that the vector and its first derivatives vanish at 
the end-points of such action.  

Associated to $W^\mu$ we may define the linearly-independent operators
\beq
\label{eq:totalW}
\frac{d_W\ }{d\tau} & := & W^\mu \frac{\partial\ }{\partial x^\mu}
+\frac{dW^\mu}{d\tau}\frac{\partial\ }{\partial \dot{x}^\mu}
+\frac{d^2 W^\mu}{d\tau^2}\frac{\partial \ }{\partial\ddot{x}^\mu}  ,\\ 
\label{eq:IW1}
I_W^{(1)} & := & W^\mu \frac{\partial\ }{\partial \ddot{x}^\mu}  ,\\
\label{eq:IW2}
I_W^{(2)} & := & W^\mu \frac{\partial\ }{\partial \dot{x}^\mu}
+2\frac{dW^\mu}{d\tau}\frac{\partial\ }{\partial \ddot{x}^\mu}  .
\eeq 
In particular, operator~(\ref{eq:totalW}) 
simply stands for the total 
derivative in the direction of the vector field $W^\mu$.  The relevance of these 
operators mainly relies on the fact that for any differentiable Lagrangian, 
$L(x,\dot{x},\ddot{x})$, 
$d_W L/d\tau$, $I_W^{(1)}(L)$ and $I_W^{(2)}(L)$ become real scalar fields,
that is, they are invariant under local coordinate transformations on $T^2 M$. 
It is straightforward to show that these operators are related to 
the Euler-Lagrange (EL) operator $E_\mu^{(0)}$, Eq.~(\ref{eq:operatorE0}),
through its projection along $W^\mu$
\beq
\label{eq:Widentity}
\frac{d_W\ }{d\tau}-\frac{d\ }{d\tau}\left( I_W^{(2)} -  
\frac{d\ }{d\tau} I_W^{(1)} \right) = W^\mu E_\mu^{(0)}  .
\eeq
Note that, for the particular case 
$W^\mu=dx^\mu/d\tau$, the operator~(\ref{eq:totalW}) becomes 
the total derivative with respect to the parameter $\tau$, while 
(\ref{eq:IW1}) and (\ref{eq:IW2}) become
identical to the Lie derivatives along the flows of the so-called 
{\it Liouville vector fields}
\beq
\Gamma^{(1)} & := & \dot{x}^\mu \frac{\partial\ }{\partial \ddot{x}^\mu},
\label{eq:Gamma1}
\\
\Gamma^{(2)} & := & \dot{x}^\mu \frac{\partial\ }{\partial \dot{x}^\mu}
+2\ddot{x}^\mu \frac{\partial\ }{\partial \ddot{x}^\mu},
\label{eq:Gamma2}
\eeq
that is, $I_{\dot{x}}^{(1)}=\pounds_{\Gamma^{(1)}}$ and 
$I_{\dot{x}}^{(2)}=\pounds_{\Gamma^{(2)}}$.  For any differential 
Lagrangian, $I_{\dot{x}}^{(1)}(L)$ and $I_{\dot{x}}^{(2)}(L)$ are called the {\it main 
invariants of the Lagrangian $L$} and, for the specific choice of our Lagrangian~(\ref{eq:lag1}), 
these become
\beq
\label{eq:invariantI1}
I_{\dot{x}}^{(1)}(L) & = & \dot{x}^\mu K_\mu  ,\\
\label{eq:invariantI2}
I_{\dot{x}}^{(2)}(L) & = & \left( 
\frac{\partial K_\mu}{\partial \dot{x}^\nu} \dot{x}^\nu + 2K_\mu \right)
\ddot{x}^\mu + \frac{\partial V}{\partial \dot{x}^\mu} \dot{x}^\mu  .
\eeq
Both invariants~(\ref{eq:invariantI1}) and~(\ref{eq:invariantI2}) serve 
to establish conservation theorems, as we will see below.
Also, from these last relations we may deduce Zermelo conditions~\cite{Kondo}, which state the 
necessary conditions for an action integral to be independent of the parametrization
of the curve $c$, namely,
\beq
I_{\dot{x}}^{(1)}(L)  =  0 ,  
\hspace{7ex}
I_{\dot{x}}^{(2)}(L)  =  L .
\label{eq:Zermelo}
\eeq
Indeed, the first of these conditions~(\ref{eq:Zermelo}) stands for the invariance 
of the Lagrangian along the vector field $I_{\dot{x}}^{(1)}$, while the second sets 
$I_{\dot{x}}^{(2)}$ as 
a genuine Liouville vector field when applied to the Lagrangian function.
In our notation, Zermelo conditions may be explicitly obtained by combining 
Eqs.~(\ref{eq:invariantI1}) to~(\ref{eq:Zermelo}).

Now, the so-called {\it energies for a second-order Lagrangian $L$} are defined
in terms of the main invariants of $L$ as
\beq
\mathcal{E}_c^{(1)}(L) & := & - I_{W}^{(1)} (L) = - W^\mu \frac{\partial L}{\partial \ddot{x}^\mu},
\nn
\\
&=& - W^\mu P_\mu,
\label{eq:GenEnergy1}
\\
\mathcal{E}_c^{(2)}(L) & := & I_{W}^{(2)}(L) - \frac{d I_{W}^{(1)}(L) }{d\tau}
- L = W^\mu \frac{\partial L}{\partial \dot{x}^\mu} + 2 \dot{W}^\mu \frac{\partial L}{\partial 
\ddot{x}^\mu} - \frac{d}{d\tau} \left( W^\mu \frac{\partial L}{\partial \ddot{x}^\mu }\right)
- L,
\nn
\\
&=& W^\mu p_\mu + \dot{W}^\mu P_\mu - L,
\label{eq:GenEnergy2}
\eeq
where we have used the definitions of the momenta~(\ref{eq:P1}) and (\ref{eq:p1}).
Conservation of these energies is dictated by the relations 
\beq
\frac{d\mathcal{E}_c^{(1)}(L)}{d\tau} & = & - \frac{1}{2} \left[ 
I_{W}^{(2)} (L) + W^\mu E_\mu^{(1)}(L)  \right],
\\
\frac{d\mathcal{E}_c^{(2)}(L)}{d\tau} & = & - W^\mu E_\mu^{(0)}(L) .
\eeq
The first of these identities may be obtained straightforwardly, 
while the second is a consequence of identity~(\ref{eq:Widentity}).
Here, we used the covector fields
$E_\mu^{(0)}$ (defined in~(\ref{eq:operatorE0})), 
$E_\mu^{(1)}:=-\partial/\partial\dot{x}^\mu+2 d/d\tau(\partial/\partial\ddot{x}^\mu)$
which, together with $E_\mu^{(2)}:=\partial/\partial\ddot{x}^\mu$,
are the so-called {\it Craig-Synge covectors} associated to a differentiable 
second-order Lagrangian~\cite{Craig,Synge}.  
When $W^\mu = \dot{x}^\mu$, it is straightforward to see that $\mathcal{E}_c^{(2)}$, 
which is related to the canonical Hamiltonian $H_0$ (see~(\ref{eq:H0}) or (\ref{eq:H00})), is conserved only along 
the solution curve to EL equations $E_\mu^{(0)}(L)= 0$.  Furthermore, 
we see from equations~(\ref{eq:GenEnergy1}) and~(\ref{eq:GenEnergy2}) that 
these energies are identically vanishing whenever Zermelo conditions~(\ref{eq:Zermelo})
are considered, that is, for covariant systems.

Finally, in order to study  the behaviour of the function~(\ref{eq:Noether}), 
we choose two points, 
$(x,\tau)$ and $(x',\tau')$ belonging to the same domain of a local chart 
$U\times (a,b)\subset M\times\mathbb{R}$.  These points are connected 
through an infinitesimal transformation of the form~(\ref{eq:infinitesimal}),
where $\epsilon\in\mathbb{R}$ is a sufficiently small positive number, and 
$\eta:=\eta(x,\tau)$ is an arbitrary smooth function locally defined 
at the point $(x,\tau)$.  Then, it is direct to show that the infinitesimal 
transformation~(\ref{eq:infinitesimal}) is a local symmetry of the Lagrangian 
$L(x,\dot{x},\ddot{x})$ if and only if for any $C^\infty$-function $F(x,\dot{x})$
the following equation holds
\beq
L\left( x',\frac{dx'}{d\tau'},\frac{d^2 x'}{d\tau'^2} \right)d\tau'
= \left[L\left( x,\frac{dx}{d\tau},\frac{d^2 x}{d\tau^2}\right)+\frac{d\ }{d\tau}
\left(F\left( x,\frac{dx}{d\tau} \right)\right)\right]d\tau  .
\eeq
From this last relation, we may Taylor expand the left hand side around unprimed
coordinates, and keeping first order terms in $\epsilon$, we may find, after some 
calculus, the identity
\beq
\frac{d\mathcal{Q}(L,\phi)}{d\tau} = \left( W^\mu-\eta \dot{x}^\mu \right)
E_\mu^{(0)}(L) ,
\eeq
where we have defined the function 
\beq
\mathcal{Q}(L,\phi):= I_W^{(2)}(L) -  \frac{d I_W^{(1)}(L) }{d\tau}
- \eta \, \mathcal{E}_c^{(2)}(L) + \dot{\eta}\, \mathcal{E}_c^{(1)}(L) - \phi ,
\label{eq:QQ}
\eeq
and $\dot{\eta} = d\eta/d \tau$ and 
$\phi$ stands for the first-order term in the 
$\epsilon$-expansion 
of the function $F(x,\dot{x})$, that is, $F(x,\dot{x})=\epsilon\phi(x,\dot{x})$.
From this, we are ready to establish the 
Noether theorem, which 
state that, along the solution curves of EL equations of motion
$E_\mu^{(0)}(L)=0$, the function $\mathcal{Q}(L,\phi)$ is conserved under 
evolution of the parameter $\tau$.  Note that the  conserved function 
$\mathcal{Q}(L,\phi)$ depends solely on the invariants 
$I_W^{(1)},\ I_W^{(2)}$ and the energies $\mathcal{E}_c^{(1)},\ \mathcal{E}_c^{(2)}$.
In physical grounds, $\mathcal{Q}$ is known as the Noether charge.
Also, we must note that whenever Zermelo conditions~(\ref{eq:Zermelo}) hold, 
the conserved function $\mathcal{Q}(L,\phi)$ is reduced to
\beq
\label{eq:Gcovariant}
\mathcal{Q}(L,\phi)= I_W^{(2)}(L) -  
\frac{d\ }{d\tau} I_W^{(1)}(L) -\phi ,
\eeq
thus, we expect~(\ref{eq:Gcovariant}) to be conserved for covariant theories.

\section{On Helmholtz conditions}
\label{appendix}

Concerning the integrability conditions for $M_{\mu \nu}$ and $K_\mu$,
our starting point will be the matrix introduced in~(\ref{eq:Mmunu}). Making use of the
partial derivatives with respect to the coordinates $x^\rho$ and 
considering the skew-symmetric part with respect to the indices 
$\mu$ and $\rho$, respectively,
we get
\begin{equation}
 \frac{\partial M_{\mu \nu}}{\partial x^\rho} - \frac{\partial 
M_{\rho \nu}}{\partial x^\mu} = \frac{\partial^2 P_\nu}{\partial x^\rho
\partial x^\mu} - \frac{\partial^2 P_\nu}{\partial x^\mu
\partial x^\rho} + \frac{\partial }{\partial \dot{x}^\nu}
\left( \frac{\partial p_\rho}{\partial x^\mu} -  \frac{\partial 
p_\mu}{\partial x^\rho} \right)   \,. 
\end{equation}
From equation~(\ref{eq:Xmunu}) we thus have
\begin{equation}
\frac{\partial X_{\mu \nu}}{\partial \dot{x}^\rho} =
\frac{\partial M_{\mu \nu}}{\partial x^\rho} - \frac{\partial 
M_{\rho \nu}}{\partial x^\mu}.
\label{eq:h-1}
\end{equation}

Now, by taking partial derivatives of the force term (\ref{eq:force}) with 
respect to $\dot{x}^\mu$, and taking the symmetric part with respect to 
the indices $\mu$ and
$\nu$, we get

\begin{equation}
\frac{\partial F_\mu}{\partial \dot{x}^\nu} + \frac{\partial F_\nu}{\partial 
\dot{x}^\mu} = \dot{x}^\alpha \frac{\partial}{\partial x^\alpha}
\left( \frac{\partial p_\mu}{\partial \dot{x}^\nu}  + \frac{\partial p_\nu}{\partial 
\dot{x}^\mu}\right) + \frac{\partial}{\partial x^\nu} \left( p_\mu -
\frac{\partial V}{\partial \dot{x}^\mu}\right) + \frac{\partial}{\partial x^\mu} 
\left( p_\nu - \frac{\partial V}{\partial \dot{x}^\nu}\right) .  
\end{equation}
Definition~(\ref{eq:p1}) can now be used to express the last two terms 
on the right hand side of this equation in terms of the highest momenta $P_\mu$
as
\begin{equation}
\frac{\partial F_\mu}{\partial \dot{x}^\nu} + \frac{\partial F_\nu}{\partial 
\dot{x}^\mu} = \dot{x}^\alpha \frac{\partial}{\partial x^\alpha}
\left( \frac{\partial p_\mu}{\partial \dot{x}^\nu}  + \frac{\partial p_\nu}{\partial 
\dot{x}^\mu}\right) - \dot{x}^\alpha \frac{\partial}{\partial x^\alpha} \left(
\frac{\partial P_\mu}{\partial {x}^\nu} + \frac{\partial P_\nu}{\partial {x}^\mu}\right). 
\end{equation}
Now, by splitting the momenta $p_\mu$ in terms of $\mathbf{p}_\mu$ and $\mathfrak{p}_\mu$,
and taking into account the identity (\ref{eq:condition-2}) we obtain
\begin{equation}
\frac{\partial F_\mu}{\partial \dot{x}^\nu} + \frac{\partial F_\nu}{\partial 
\dot{x}^\mu} = \dot{x}^\alpha \frac{\partial}{\partial x^\alpha}
\left( \frac{\partial \mathbf{p}_\mu}{\partial \dot{x}^\nu}  + \frac{\partial 
\mathbf{p}_\nu}{\partial \dot{x}^\mu}\right) ,
\end{equation}
whereby we will have, from equation~(\ref{eq:Mmunu-2}), the condition
\begin{equation}
2 \dot{x}^\alpha \frac{\partial M_{\mu \nu}}{\partial x^\alpha}
= - \left( \frac{\partial F_\mu}{\partial \dot{x}^\nu} + \frac{\partial F_\nu}{\partial 
\dot{x}^\mu} \right). 
\label{eq:h-2}
\end{equation}

Similarly as for (\ref{eq:h-2}), 
by taking partial derivatives 
of~(\ref{eq:force}) with respect to ${x}^\mu$, and 
considering the symmetric part  
with respect to indices $\mu$ and $\nu$, we get
\begin{equation}
\dot{x}^\alpha \frac{\partial X_{\mu \nu}}{\partial x^\alpha}
= \frac{\partial F_\nu}{\partial {x}^\mu} + \frac{\partial F_\mu}{\partial x^\nu} .
\label{eq:h-3}
\end{equation}
Expressions~(\ref{eq:Mmunu-2}),~(\ref{eq:M-1}),~(\ref{eq:h-2})
and~(\ref{eq:h-3}) are usually referred to as the Helmholtz integrability
conditions satisfied for a non-singular matrix $M_{\mu \nu}$ and 
the vector $K_\mu$ in order to obtain eom~(\ref{eq:eom2}) from a variational principle.



\begin{thebibliography}{99}

\bibitem{hojman1} 
R. Hojman, S. Hojman and J. Sheinbaum,
Phys.~Rev.~D {\bf 28}, 1333 (1983). 

\bibitem{hojman2} 
S. Hojman, 
J. Phys. A {\bf 17}, 2399 (1984).

\bibitem{hojman3} 
R. Hojman and J. Zanelli, 
Phys. Rev. D {\bf 35}, 3825 (1987).

\bibitem{Foster}
B. L. Foster, 
Proc.~Royal Soc.~London A: Math.~Phys.~Sci. {\bf 423}, 443 (1989).

\bibitem{udriste} 
C. Udri\c{s}te and A. Pitea, 
Balkan Journal of Geometry and its Applications {\bf 16}, 174--185 (2011).

\bibitem{barker}
B. M. Barker and R. F. O'Connell, 
Phys.~Lett. A {\bf 78}, 231--232 (1980).

\bibitem{tesser}
H. Tesser, 
J.~Math.~Phys. {\bf 13}, 796--799 (1972).

\bibitem{popescu1} 
P. Popescu, 
J.~Geom.~Phys. {\bf 77}, 113 (2014); arXiv: 1212.4873v2 [math-ph]

\bibitem{popescu2} 
P. Popescu and M. Popescu, 
Balkan J.~Geom.~Appl. {\bf 17}, 82--91 (2012).

\bibitem{Pirani}
F. A. E. Pirani and A. Schild, 
Phys.~Rev. {\bf 79}, 986 (1950).

\bibitem{ostro-prd} 
R. Cordero, A. Molgado and E. Rojas, 
Phys. Rev. D {\bf 79}, 024024 (2009).

\bibitem{membrane} 
R. Cordero, A. Molgado and E. Rojas, 
Class. Quant. Grav. {\bf 28}, 065010 (2011).

\bibitem{mgbg} 
R. Cordero, M. Cruz, A. Molgado and E. Rojas, 
Class. Quant. Grav. {\bf 29}, 175010 (2012). 

\bibitem{ccmr}
R. Cordero, M. Cruz, A. Molgado and E. Rojas, 
Gen.~Rel.~Grav. {\bf 46}, 1761 (2014); 

\bibitem{Dutt}
S. K. Dutt and M. Dresden, 
Preprint ITP-SB-86-32 Stony Brook, (1986).

\bibitem{Ghalati}
R. N. Ghalati, N. Kiriushcheva and S. V. Kuzmin, 
Mod.~Phys.~Lett. A {\bf 22}, 17--28 (2007). arXiv: hep-th/0605193v3

\bibitem{Paul}
B. Paul, 
Phys.~Rev.~D ~{\bf 87}, 045003 (2013); arXiv: 1212.5902v2 [hep-th].

\bibitem{PaulRT}
R. Banerjee, P. Mukherjee and B. Paul, 
Phys.~Rev. D {\bf 89}, 043508 (2014); arXiv: 1307.4920v2 [gr-qc].

\bibitem{hojman4} 
R. Hojman and J. Zanelli, 
Nuovo Cimento {\bf 94}, 87 (1986).

\bibitem{deleon}
M. De Le\'on and J. C. Marrero, Diff.~Geom.~Appl. Proc.~Conf.~Opava 497--508, (1993);

\bibitem{pardo} 
F. Pardo, 
J.~Math.~Phys.~{\bf 30}, 2854 (1989).

\bibitem{Gonera-note}
K. Andrzejewski, J. Gonera and P. Ma\'slanka, \textit{A note on the Hamiltonian formalism for 
higher-derivative theories}, arXiv: 0710.2976v1 [hep-th].

\bibitem{Kondo}
K. Kondo, 
Tensor N.~S. {\bf 14}, 191--215 (1963).


\bibitem{MironL}
R. Miron, \textit{The Geometry of Higher-Order Lagrange spaces: Applications to Mechanics and 
Physics} (Kluwer Academic Publishers, 1997)

\bibitem{MironH}
R. Miron, {\it The Geometry of Higher-Order Hamilton spaces: Applications to Hamiltonian Mechanics}
(Kluwer Academic Publishers, 2003)

\bibitem{MironNoether}
R. Miron, {\it Int.~J.~Theor.~Phys.} {\bf 34} 1123--1146 (1995)

\bibitem{matsyuk}
R. Ya Matsyuk, {\it J.~Diff.~Geom.~Appl.}
{\bf  29}  S149--S155 (2011), \textit{arXiv:}1101.5384v2 [math.DG]

\bibitem{gracia}
X. Gr\`acia, J. M. Pons and N. Rom\'an-Roy,
\textit{J.~Phys.~A: Math.~Gen.} {\bf 25}, (1989)

\bibitem{gracia2}
X. Gr\`acia, J. M. Pons and N. Rom\'an-Roy,
\textit{J.~Math.~Phys.} {\bf 32}, 2744 (1992)

\bibitem{dediego}
M. de Le\'on and D. Mart\'in de Diego,
\textit{J.~Math.~Phys.} {\bf 36}, 4138 (1995)

\bibitem{capo95}
R. Capovilla and J. Guven, \textit{Phys. Rev. D} \textbf{51} 6736-6743 (1995). 





\bibitem{Govaerts} 
J. Govaerts, 
Int. J. Mod. Phys. A {\bf 5}, 3625--3640 (1990).

\bibitem{exorcist}
T-j. Chen, M. Fasiello, E. A. Lim and A. J. Tolley, 
J. Cosmol. Astrop. Phys. {\bf 02}, 042 (2013); arXiv: 1209.0583v4 [hep-th]. 

\bibitem{nesterenko-insta}
V. V. Nesterenko, 
Phys.~Rev. D {\bf 75}, 087703 (2007); arXiv: hep-th/0612265v2.

\bibitem{Stephen}
N. G. Stephen, 
J.~Sound Vib. {\bf 310}, 729--739 (2008).

\bibitem{llosa}
J. Llosa, 
Phys.~Rev. A {\bf 67}, 016101 (2003); arXiv: hep-th/0201087v1.

\bibitem{querella} 
L. Querella, Ph.D. Thesis, 
Universit\'e de Li\'ege, Facult\'C des sciences, 1998; arXiv: gr-qc/9902044v1.

\bibitem{davidson}
A. Davidson and D. Karasik, Mod. Phys. Lett. A {\bf 13}, 2187--2192 (1998);
Phys. Rev. D {\bf 67}, 064012 (2003).



\bibitem{ostro} 
M. Ostrogradski, Mem.~Ac.~St.~Petersbg. {\bf V1}, 385 (1850).

\bibitem{nesterenko1} 
V. V. Nesterenko, 
J. Phys. A {\bf 22}, 1673 (1989).

\bibitem{Dirac}
P. A. M. Dirac, {\it Lectures on Quantum Mechanics} (Dover publications, Mineola, New York, 2001).

\bibitem{gauge} 
M. Henneaux and C. Teitelboim, {\it Quantization of Gauge Systems} (Princeton University Press, Princeton, 
New Jersey, 1992).

\bibitem{rothe} 
H. Rothe and K. Rothe, {\it Classical and Quantum Dynamics of Constrained Hamiltonian Systems}
(World Scientific Lectures Notes in Physics, Vol. 81, 2010).

\bibitem{nesterenko2} 
V. V. Nesterenko,
Phys.~Lett.~B~{\bf 327}, 50 (1994).

\bibitem{sarlet} 
W. Sarlet, 
J.~Phys.~A~{\bf 15}, 1503 (1982).

\bibitem{henneaux2} 
M. Henneaux, 
Ann.~Phys.~{\bf 140}, 45 (1982).

\bibitem{lukierski} 
J. Lukierski, P. Stichel and W. Zakrzewski, 
Ann.~Phys. {\bf 260}, 224 (1997); arXiv: hep-th/9612017v2.

\bibitem{Gonera1}
K. Andrzejewski, J. Gonera, P. Machalski and P. Ma\'slanka, 
Phys.~Rev. D {\bf 82}, 045008 (2010); arXiv: 1005.3941v4 [hep-th].

\bibitem{Gonera2}
P. M. Zhang, P. A. Horvathy, K. Andrzejewski, J. Gonera and P. Kosi\'nski, 
Ann.~Phys. {\bf 333}, 335--359 (2012); arXiv: 1207.2875v3 [hep-th].

\bibitem{Acatrinei}
C. S. Acatrinei, 
J.~Phys.~A:~Math.~Theor.~{\bf 40}, F929--F934 (2007); arXiv: 0708.4351 [hep-th].

\bibitem{Horvathy}
P. A. Horvathy and M. S. Plyushchay, 
J.~High~Energy~Phys.~{\bf 0206}, 033 (2002); arXiv: hep-th/0201228.

\bibitem{2ndorder}
C. Udri\c{s}te, 
in Proceedings of the Workshop on Global Analysis, Differential Geometry and Lie Algebras 1998, 
161--168, Balkan Society of Geometers (1999).

\bibitem{otsuki} 
T. Otsuki, 
Math.~J.~Okayama Univ.~{\bf 32}, 227--242 (1990).

\bibitem{Tu} 
P. N. V. Tu, {\it Introductory Optimization Dynamics} (Springer-Verlag, 1991).

\bibitem{leon-rds}
M. De Le\'on and P. R. Rodrigues, {\it Generalized Classical Mechanics 
and Field Theory: A geometrical approach of Lagrangian and Hamiltonian 
Formalisms involving Higher Order derivatives} (Elsevier Science Publishers, 1985).

\bibitem{hambranes} 
R. Capovilla, J. Guven and E. Rojas, 
Class.~Quant.~Grav.~{\bf 21} 5563--5586 (2004); arXiv: hep-th/0404178

\bibitem{note}
In any $N$-dimensional manifold with a non-degenerate 
metric $g_{\mu \nu}$ that is used to lower and raise indexes, for covariant 
theories defined in an immersed $(p+1)$-dimensional surface, at any point of 
the surface the following decomposition holds
\begin{equation}
\nn
g^\mu {}_\nu = {\cal H}^\mu {}_\nu + \perp^\mu{}_\nu ,
\label{eq:projector}
\end{equation}
where ${\cal H}^{\mu \nu}= g^{ab} e^\mu{}_a e^\nu{}_b$ is the projection
tensor of rank $(p+1)$ on the surface, $g_{ab}$ is the induced metric, and 
$e^\mu{}_a$ denotes the tangent vectors to the surface. Further, $\perp^{\mu\nu}
= n^\mu{}_{(s)} n^{\nu} _{(s)}$ is the complementary projector tensor of rank $(N-p-1)$,
orthogonal to the surface.

\bibitem{onder} 
M. \"{O}nder and R. M. Tucker, 
J.~Phys.~A.~Math.~Gen. {\bf 21}, 3423 (1988)

\bibitem{Craig}
H. V. Craig,
Am.~J.~Math. {\bf 57}, 457--462 (1935).

\bibitem{Synge}
J. L. Synge, 
Am.~J.~Math. {\bf 57}, 679--691 (1935).

\end{thebibliography}
\end{document}